\documentclass{IEEEtran}
\usepackage{cite}
\usepackage{amsmath,amssymb,amsfonts}
\usepackage{algorithm}
\usepackage{algpseudocode}
\usepackage{textcomp}
\usepackage[caption=false,font=normalsize,labelfont=sf,textfont=sf]{subfig}
\usepackage{stfloats}
\usepackage{url}
\usepackage{verbatim}
\usepackage{hyperref}
\usepackage{tikz}
\usepackage{balance}
\usepackage{booktabs}
\usepackage{multirow}
\usetikzlibrary{arrows.meta}
 \usepackage{pgfplots}
 \pgfplotsset{compat=1.18}

\newtheorem{lemma}{Lemma}

\def\BibTeX{{\rm B\kern-.05em{\sc i\kern-.025em b}\kern-.08em
    T\kern-.1667em\lower.7ex\hbox{E}\kern-.125emX}}
\begin{document}

\title{Passive Beam Shaping via Binary-Coded Apertures}


\author{Mohammed E. Eltayeb
\thanks{This work was supported by the National Science Foundation
under Grant No. NSF-2243089. }
\thanks{Mohammed E. Eltayeb is with the Department of Electrical and Electronic Engineering,  California State University, Sacramento, USA.
(Email: mohammed.eltayeb@csus.edu) }
}

\maketitle

\begin{abstract}
 This paper presents coded-aperture reflector for indoor   {millimeter-wave (mmWave)} coverage enhancement in obstructed or blocked   {line-of-sight (LoS)} settings.We model the reflecting aperture using an equivalent array-factor formulation, where each passive reflecting cell contributes a reradiated field with phase set by the incident and departure directions.  Building on this model, we develop two fabrication-friendly passive synthesis methods: (i) {binary (1-bit) spatial coding} that enables deterministic non-specular beam formation and multi-beam patterns by selecting cell participation on a dense lattice via an ON/OFF metallization mask, and (ii) {diffraction-order (periodic) steering} that exploits aperture periodicity to place selected diffraction orders at prescribed angles. We analytically characterize the proposed {cosine-threshold} quantization rule,  including its asymptotic activation ratio and a distribution-free lower bound on non-specular gain relative to ideal continuous-phase control. To validate the proposed designs, we fabricate and metallize low-cost prototypes {in-house} using a copper-backed 3D-printed ``inkwell'' substrate with stencil-guided conductive-ink deposition. $60$~GHz over-the-air measurements show non-specular gain enhancements on the order of $+14$--$20$~dB relative to passive, non-engineered (all-ON) reflector baselines.  Results also demonstrate that fully passive,  binary-coded apertures can deliver beam control with rapid in-lab manufacturability and offer a practical alternative to power-consuming reconfigurable surfaces for static indoor mmWave links.
\end{abstract}


\section{Introduction}
\label{sec:introduction}
{M}{illimeter-wave} (mmWave, $30$--$300\,\mathrm{GHz}$) and terahertz (THz, $0.1$--$10\,\mathrm{THz}$)
bands enable multi-Gbps wireless links via wide contiguous bandwidths and support high-angular-resolution operation
through highly directive apertures \cite{wang2023on6g,xue2024beam,ahmed2024ris}.
At the same time, short wavelengths intensify propagation challenges, including higher free-space loss,
increased penetration loss through common building materials, and pronounced sensitivity to blockages and
geometric shadowing \cite{xing2021millimeter,heath2016overview}.
Consequently, practical indoor mmWave links often exhibit \emph{persistent} coverage holes in obstructed-/blocked-LoS
geometries such as corridor turns, non-specular corners, and junctions (e.g., L- and T-intersections), where even a
strongly illuminated wall cannot deliver energy toward the user through mirror-like specular reflection.
This motivates \emph{electromagnetic environment shaping}, in which reflections are deliberately engineered to create reliable redirected paths that extend connectivity into   {non-line-of-sight (NLoS)} regions without requiring additional active infrastructure.

A prominent environment-shaping approach is the reconfigurable intelligent surface or large intellgent surface (RIS or LIS), which employs tunable unit cells (e.g., varactors,    {positive-intrinsic-negative (PIN) diodes, microelectromechanical systems (MEMS), }or liquid crystals) to control the reflection coefficient across
an aperture and redirect incident wavefronts \cite{ahmed2024ris,HuangRIS2019,khateeb_LiS,elmossallamy2020reconfigurable,hum2014reconfigurable}.
RIS can provide adaptivity, multi-user beam management, and programmable wavefront control, and has therefore
received substantial attention in the communications and antennas communities. However, practical RIS deployments
must contend with several non-idealities and overheads that are often overlooked in idealized phase-only models.
First, finite-resolution phase tuning (quantization), mutual coupling, and element loss can reduce the realized
array gain and distort sidelobe structure, especially at mmWave where fabrication tolerances and parasitics matter
\cite{hum2014reconfigurable,elmossallamy2020reconfigurable}. Second, RIS operation typically relies on a control
loop (CSI acquisition, optimization, and configuration), which can impose signaling overhead and latency, and can
become challenging when the channel changes or when users are mobile \cite{LiuComST2021_RISPrinciplesOpp}.
Third,  and central to the deployment perspective of this work,  practical RIS require non-negligible DC power for
controllers, driver/bias circuitry, and tunable elements, even if the RF scattering mechanism itself does not
include a transmit chain \cite{HuangRIS2019,WangTCOM2024Power,WangEUSIPCO2023StaticPower}. Measurement-backed studies
show that controller/driver power can be at the watt level and that unit-cell dissipation can be state dependent
(e.g., for PIN-diode RIS, ON-state cells consume additional power) \cite{WangTCOM2024Power,WangEUSIPCO2023StaticPower}.
For static indoor installations, these operational burdens translate into wiring complexity, power provisioning,
maintenance considerations, and cost factors that matter in real deployments but are not captured by the usual
``nearly-passive'' narrative.

To reduce tuning and control complexity, recent work has explored quantized reconfigurable architectures
\cite{Gros2021OJCOMS,Kamoda2011TAP,Shekhawat2025OJAP,Naqvi2023Micromachines,Huang2023SciRep,li2021onebit}.
Binary-phase RIS/metasurfaces using PIN diodes have been demonstrated at mmWave frequencies \cite{Gros2021OJCOMS},
and single-bit phase-shifter reflectarrays have been used for 60\,GHz electronic steering \cite{Kamoda2011TAP}.
Quantization artifacts such as grating/quantization lobes can be mitigated using additional structure (e.g.,
pseudo-random fixed delays combined with $0/\pi$ switching) \cite{Shekhawat2025OJAP}. While these platforms
simplify phase control relative to multi-bit RIS, they still require bias networks, scalable routing/control, and
continuous surface-side power whose overhead increases with aperture size \cite{li2021onebit,WangTCOM2024Power}.
Thus, there remains a practical gap between (i) simple passive metallic reflectors that are robust and power-free
but largely constrained to specular reflection, and (ii) fully programmable RIS that are flexible but impose
nontrivial operational overhead.

This gap has renewed interest in \emph{fully passive} reflectors for static or quasi-static deployments,
where robustness, simplicity, and zero operational power are desirable. Prior studies show that simple passive
reflectors (e.g., metallic panels) can enhance mmWave coverage in NLoS settings by creating strong redirected
paths when placed appropriately \cite{khawaja2020coverage,peng2016,ganesh2023propagation,eltayeb2025lidar}.
Beyond canonical plates, custom passive surfaces and application-specific passive designs have been explored to
shape reflections more deliberately \cite{hager2024custom,qian2022millimirror,qian2022fully}. A representative
example is \emph{MilliMirror} \cite{qian2022millimirror,qian2022fully}, which encodes a desired reflection phase
profile through {geometry} (height) modulation of metal-backed dielectric structures. Geometry-coded phase
panels can be highly effective, but they typically require a custom 3D thickness profile per beam pattern objective and
can impose tighter tolerance requirements (e.g., printer $z$-resolution, warping/shrinkage, and material
permittivity variation) when scaling to larger apertures.

\begin{figure}[t]
\centering
\includegraphics[width=0.30\textwidth]{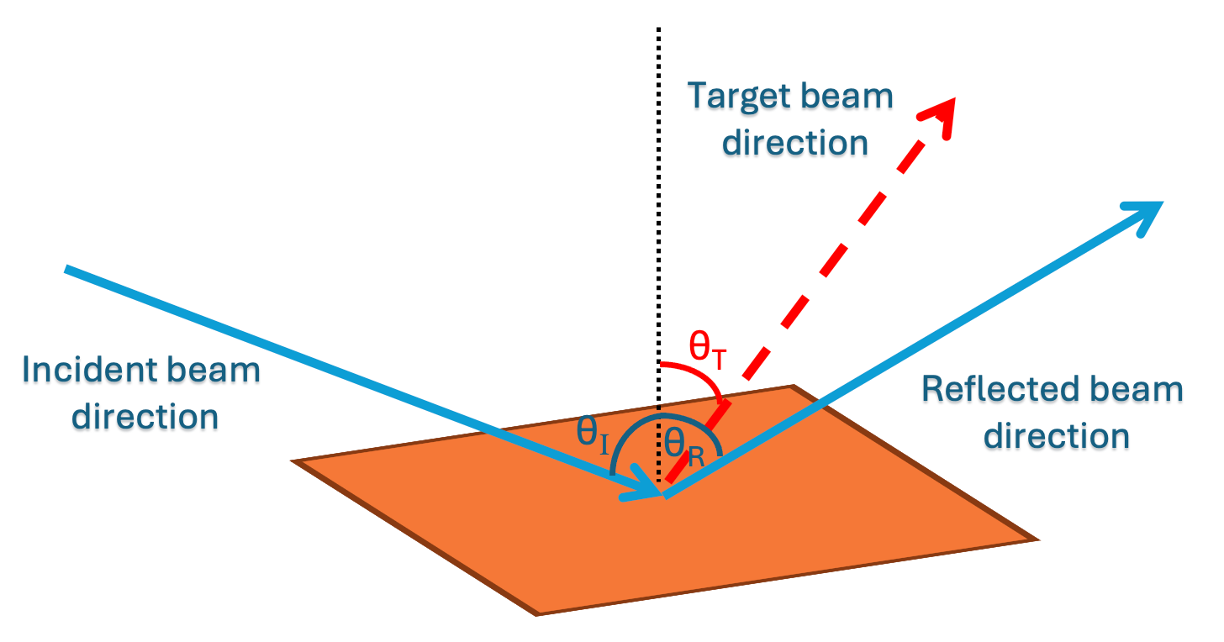}
\caption{Equivalent array-factor model for a passive reflecting element. Each passive element is modeled as an
equivalent radiator whose far-field contribution carries the tangential phase associated with the incident and
reradiated directions.}
\label{fig:model}
\end{figure}

  {A practical gap remains between simple passive reflectors, which are robust and power-free but largely limited to specular redirection, and reconfigurable surfaces, which provide flexible beam control at the cost of active hardware, control circuitry, and operational power. Geometry-coded passive reflectors offer another alternative, but they typically require a new 3-D design for each beam objective. To address this gap, this paper proposes a novel copper-backed 3D-printed inkwell reflector platform that combines a fixed dense scaffold with analytically derived binary metallization patterns.  This structure enables rapid realization of different beam patterns on a common base reflector design, without requiring a new 3-D reflector geometry for each target response.

Within this framework, this work} focuses on \emph{binary-coded apertures} that generate deterministic non-specular beams using simple passive coding patterns. The reflecting aperture is modeled using an equivalent array-factor formulation, in which each passive reflecting cell contributes a reradiated field whose phase is determined by the incident and departure directions, as shown in Fig.~\ref{fig:model}. Building on this model, two fabrication-friendly passive synthesis methods are developed. First, a closed-form \emph{cosine-threshold} rule is proposed to convert the ideal continuous phase ramp for one or more target directions into a fabrication-ready ON/OFF metallization mask on a dense $\lambda/2$ lattice. This \emph{binary (1-bit) spatial coding} enables beam steering and multi-beam formation by selecting which cells participate in coherent addition. Second, \emph{diffraction-order (periodic) steering} is developed by exploiting controlled aperture periodicity to place selected diffraction orders at prescribed angles. Both methods are strictly passive and require no bias network, tuning circuitry, or surface-side power.

The proposed coded apertures are best suited for quasi-static indoor settings where the geometry is known (or changes slowly) and where specular reflection alone does not illuminate the intended region.  Representative cases include: (i) \emph{L-corners and hallway turns} (including non-right-angle turns and gently curved corridors), where the dominant {specular} (mirror-angle) reflection fails to illuminate the region beyond the turn and  coverage extension requires an engineered {non-specular} departure lobe toward the user; (ii) \emph{T-intersections and corridor junctions}, where energy must be redirected into a side branch outside the specular footprint; and (iii) \emph{doorway and room-corridor transitions}, where a panel near an opening must steer energy through the doorway to reach the interior.  The approach is also useful in indoor multipath environments, where the propagation geometry supports multiple distinct reflected rays from the {same} panel reaching the {same} user location. In such settings, a dominant specular component may be present but can be intermittently blocked or degraded by local destructive interference. By synthesizing one or more deterministic non-specular departure lobes, the reflector intentionally creates {reflector-induced multipath} to the same user. This improves robustness without reflector-side power requirement.  

To validate the proposed theory-to-hardware pipeline, prototypes are fabricated using a copper-backed 3D-printed ``inkwell'' substrate with stencil-guided conductive-ink deposition.    {The reflector realization combines a fixed inkwell scaffold, an analytically derived binary ON/OFF mask pattern, and a stencil used to deposit conductive ink only into the selected wells. This enables rapid prototyping, since identical inkwell bases can be prepared in advance and metallized according to different mask patterns. } Over-the-air (OTA) measurements at 60.48~GHz in single- and multi-beam configurations closely match the predicted steering behavior, confirming that fully passive binary-coded apertures can deliver deterministic non-specular beam formation with essentially zero operational power and deployment-friendly manufacturability.

The main contributions of this paper are summarized as follows
\begin{itemize}
\item We develop an array-factor model for binary-coded passive apertures and derive a closed-form cosine-threshold rule that maps an ideal phase profile to a fabrication-ready ON/OFF metallization mask.

\item We analytically characterize the cosine-threshold mask, including its asymptotic activation ratio and a distribution-free lower bound on achievable non-specular target-lobe gain.

\item We present a fully passive diffraction-order (periodic) design method and derive a closed-form period-selection rule to place a selected diffraction order at a prescribed departure angle.

\item We show that deterministic non-specular beam formation is achievable without per-element geometry optimization by fabrication-coding patterns on a fixed dense lattice, in contrast to height-modulated reflectors (e.g., MilliMirror \cite{qian2022millimirror,qian2022fully}).

\item We fabricate and metallize low-cost prototypes \emph{in-house} using a copper-backed 3D-printed inkwell scaffold with stencil-guided conductive-paint deposition, enabling rapid iteration by swapping masks rather than reprinting a full 3D phase surface.

\item We validate single- and multi-beam operation through $60.48$~GHz over-the-air measurements, showing enhancements at intended non-specular directions consistent with theory for both mask-coded and diffraction-order designs.
\end{itemize}
Together, these contributions establish a low-complexity, fabrication-friendly framework for fully passive reflected-field control to enhance indoor NLoS coverage.

\section{System Model and Problem Formulation}
\label{sec:model}

\subsection{System Model} {  {
We consider a narrowband mmWave link in which a transmitting node illuminates a planar,
\emph{fully passive} reflecting aperture that reradiates energy into the reflection
half-space. The fabricated reflector used in this work is a 2-D lattice of candidate cells.
Accordingly, we begin with a general 2-D planar aperture model, and then specialize it to
the azimuth-plane case considered in this paper. 

Let the reflector consist of $M_x \times M_y$ electrically reflecting cells located at }}
\begin{equation}
x_m=\left(\frac{M_x-1}{2}-m\right)d_x,\qquad m=0,1,\ldots,M_x-1,
\end{equation}
\begin{equation}
y_n=\left(\frac{M_y-1}{2}-n\right)d_y,\qquad n=0,1,\ldots,M_y-1,
\end{equation}   {
where $d_x$ and $d_y$ denote the lattice spacings along the two aperture axes. A dominant
plane-wave component impinges on the reflector from direction $(\theta_I,\phi_I)$, where
$\theta_I$ and $\phi_I$ denote the polar and azimuth angles, respectively, and the scattered
field is observed toward direction $(\theta,\phi)$. Define the directional-cosine components}
\begin{equation}
u(\theta,\phi)\triangleq \sin\theta\cos\phi,\qquad
v(\theta,\phi)\triangleq \sin\theta\sin\phi .
\label{eq:uv_def}
\end{equation}  {
Under the standard far-field array-factor abstraction commonly used for RIS/reflectarrays}
\cite{khateeb_LiS,HuangRIS2019,trees}, the complex baseband response can be written as
\begin{equation}
p(\theta,\phi;\theta_I,\phi_I)=
\sum_{m=0}^{M_x-1}\sum_{n=0}^{M_y-1}
\psi_{m,n}
e^{-jk\left[x_m(u+u_I)+y_n(v+v_I)\right]},
\label{eq:p_2d}
\end{equation}
where $k=2\pi/\lambda$, $\psi_{m,n}$ denotes the complex scattering coefficient of the
$(m,n)$th cell, and
\[
u=u(\theta,\phi),\quad v=v(\theta,\phi),\quad
u_I=u(\theta_I,\phi_I),\quad v_I=v(\theta_I,\phi_I).
\]
  {
In this paper, we are primarily concerned with beam shaping in the azimuth plane.
Accordingly, the azimuth angles are fixed to zero, i.e., $\phi=\phi_I=0$. Under this
restriction, }\eqref{eq:p_2d} reduces to an azimuth-plane cut of the 2-D aperture response,
namely
\begin{equation}
p(\theta,0;\theta_I,0)
=
\sum_{m=0}^{M_x-1}\sum_{n=0}^{M_y-1}
\psi_{m,n}\,
e^{-jkx_m\left(\sin\theta+\sin\theta_I\right)}.
\label{eq:p_2d_cut}
\end{equation}  {
Since the phase progression in} \eqref{eq:p_2d_cut}   {depends only on the $x$-coordinates, the
2-D response reduces to an equivalent one-dimensional (1-D) azimuthal model.  } For analytical
tractability, we therefore adopt the 1-D formulation along the $x$-axis with $M$
electrically reflecting cells (elements) located at
\begin{equation}
x_m=\left(\frac{M-1}{2}-m\right)d_0,\qquad m=0,1,\ldots,M-1,
\label{eq:xm_sys}
\end{equation}
where $d_0$ is the nominal lattice spacing. A dominant plane-wave component impinges on the
reflector from the angle of arrival (AoA) $\theta_I$ (measured from aperture broadside), and
the scattered field is observed toward an angle of departure (AoD) $\theta$.

Under the same far-field array-factor abstraction,  and omitting explicit zero-based indexing for clarity,  the complex baseband response can be
written in matrix form as
\begin{equation}
p(\theta,\theta_I)\triangleq \mathbf{a}^{\mathsf{T}}(\theta)\,\mathbf{\Psi}\,\mathbf{a}(\theta_I),
\label{eq:p_sys}
\end{equation}
where $\mathbf{a}(\theta)\in\mathbb{C}^{M}$ is the 1-D steering vector
\begin{equation}
\mathbf{a}(\theta)\triangleq
\left[e^{-jkx_0\sin\theta},\,e^{-jkx_1\sin\theta},\,\ldots,\,e^{-jkx_{M-1}\sin\theta}\right]^{\mathsf{T}},
\label{eq:a_sys}
\end{equation}
and $\mathbf{\Psi}=\mathrm{diag}(\psi_0,\ldots,\psi_{M-1})$ captures the element-wise complex
scattering coefficients. Expanding \eqref{eq:p_sys} yields the phasor summation
\begin{equation}
p(\theta,\theta_I)=\sum_{m=0}^{M-1}\psi_m\,e^{-jkx_m\left(\sin\theta+\sin\theta_I\right)}.
\label{eq:p_sum}
\end{equation}
A uniform passive reflector corresponds to $\psi_m=1$, which produces a dominant specular
response at $\theta_{\mathrm{spec}}=-\theta_I$.

\subsection{Problem Formulation}    {
This paper targets blocked-/obstructed-LoS indoor geometries in which extending coverage
requires an engineered non-specular reflected path. In the most general 2-D setting, given
incident direction $(\theta_I,\phi_I)$,  the design objective is to choose a strictly passive
aperture configuration that increases the response toward one or more desired departure
directions $\{(\theta_{T,\ell},\phi_{T,\ell})\}_{\ell=1}^{L}$. Equivalently, we seek passive
design variables embedded in $\{\psi_{m,n}\}$ that shape the angular power pattern
$|p(\theta,\phi;\theta_I,\phi_I)|^2$ to concentrate energy at the target direction(s).

In this paper, however, we focus on azimuth-plane beam shaping with $\phi_I=\phi=0$.
Under this restriction, the design problem reduces to choosing passive design variables
embedded in $\{\psi_m\}$ that shape the azimuthal power pattern $|p(\theta,\theta_I)|^2$ to
concentrate energy at one or more desired departure angles }
$\{\theta_{T,\ell}\}_{\ell=1}^{L}$. Using the scattering response in \eqref{eq:p_sum}, we
seek passive design parameters that shape the angular power pattern
$|p(\theta,\theta_I)|^{2}$ such that 
\begin{equation}
\begin{aligned} \nonumber
\theta \in \{\theta_{T,\ell}\}_{\ell=1}^{L}
&\Rightarrow |p(\theta,\theta_I)|^{2}\ \text{is maximized},\\
\theta \notin \{\theta_{T,\ell}\}_{\ell=1}^{L}
&\Rightarrow |p(\theta,\theta_I)|^{2}\ \text{is minimized (or bounded)}.
\end{aligned}
\label{eq:objective_generic}
\end{equation}
We pursue this goal using two complementary fully passive mechanisms:

\subsubsection{Fixed-aperture 1-bit ON/OFF spatial masking}
We realize beam shaping on a fixed dense lattice by selecting which cells participate in
\eqref{eq:p_sum} using an ON/OFF metallization mask
\begin{equation}
\psi_m=b_m,\qquad b_m\in\{0,1\}.
\label{eq:binaryPsi}
\end{equation}
Substituting \eqref{eq:binaryPsi} into \eqref{eq:p_sum} yields
\begin{equation}
p(\theta,\theta_I)=\sum_{m=0}^{M-1} b_m\,
e^{-jkx_m(\sin\theta+\sin\theta_I)},
\label{eq:p_onoff}
\end{equation}
where the binary sequence $\{b_m\}$ gates the element phasors and reshapes the spatial
spectrum of the aperture. 
\subsubsection{Diffraction-order (periodic) steering via uniform period selection}
As a second fully passive mechanism, we exploit aperture periodicity by enforcing a uniform
{effective inter-element period} $\delta$ along the steering axis. Starting from the
dense candidate lattice with spacing $d_0$, we periodically activate elements so that adjacent
{active} cells are separated by $\delta$ (typically $\delta=\kappa d_0\ge d_0$ for some integer
$\kappa$). For ideal passive elements ($\psi_m=1$ on active sites), the periodicity produces
discrete diffraction orders (grating lobes) whose angles are wavelength dependent. By choosing
$\delta$, a selected order can be aligned with a target AoD. This approach is attractive when aperture
expansion is permissible, but becomes less suitable under fixed-footprint constraints.

\subsection{Narrowband Assumption and Wideband Considerations}
\label{subsec:wideband_note} 
  {The analytical development in} \eqref{eq:p_sum}   {assumes a narrowband model. Over wide
bandwidths, the phase progression scales with $k(f)=2\pi/\lambda(f)$, which can cause
frequency-dependent beam pointing (beam squint) and target-direction gain variation when the
physical reflector geometry (mask/period) is held fixed. Because the physical cell spacing is
fixed, its electrical spacing varies with frequency; as frequency increases, the lattice becomes
electrically sparser relative to the wavelength, whereas at lower frequencies it becomes
electrically denser. This effect is typically more pronounced for diffraction-order designs
because their lobe angles satisfy a wavelength-dependent grating condition. }We quantify wideband
behavior in Section~\ref{sec:results} by sweeping frequency while keeping the physical design fixed.

\section{Passive Beam Shaping via Fixed-Aperture 1-Bit Spatial Mask}
\label{sec:problem_mask}
This section develops a fully passive beam-shaping method based on a fixed-aperture
1-bit {ON/OFF} spatial mask implemented on a dense lattice. The reflector is
fabrication-coded where each lattice location is either metallized (\emph{ON}) and
contributes a strong scattered field, or left unmetallized (\emph{OFF}) and is
suppressed. Under the array-factor abstraction, the ON/OFF pattern gates the
phasor sum and reshapes the angular scattering response.  The proosed synthesis strategy is as follows: (i) form the ideal continuous phase profile that yields coherent addition toward one or more desired departure directions, and (ii) map that profile to a fabrication-friendly binary mask using a closed-form cosine-threshold rule.

\subsection{Binary Mask Model}
\label{subsec:binary_mask_model}
Consider a 1-D aperture with $M$ candidate locations along the $x$-axis, spaced by
$d_0$ at positions $x_m$ in \eqref{eq:xm_sys}. Let $b_m\in\{0,1\}$ denote the state
of the $m$th location, where $b_m=1$ corresponds to an ON (reflecting) cell and
$b_m=0$ corresponds to an OFF (suppressed) cell. Define the diagonal mask matrix
\begin{equation}
\mathbf{B}\triangleq \mathrm{diag}(b_0,b_1,\ldots,b_{M-1}).
\label{eq:B_def}
\end{equation}
For an incident plane wave arriving from AoA $\theta_I$, the complex response
observed toward AoD $\theta$ is
\begin{equation}
p(\theta,\theta_I)
=\mathbf{a}^{\mathsf{T}}(\theta)\,\mathbf{B}\,\mathbf{a}(\theta_I)
=\sum_{m=0}^{M-1} b_m\,e^{-jkx_m(\sin\theta+\sin\theta_I)},
\label{eq:pb_sum}
\end{equation}
which is the specialization of \eqref{eq:p_sys} with $[\mathbf{\Psi}]_{m,m}=b_m$.
Equation~\eqref{eq:pb_sum} shows that the binary mask $\{b_m\}$ gates the element
phasors where ON cells participate in coherent addition and OFF cells are removed
from the sum.  

In practice, reflection imperfections and ohmic/dielectric loss can be
captured by replacing the ON state with an amplitude coefficient $b_m=\rho$ (with
$0<\rho\le 1$), while retaining $b_m=0$ for the OFF state. For ease of exposition,  this
paper assumes $b_m\in\{0,1\}$ and does not explicitly model such non-ideal reflection
amplitudes.

\subsection{Closed-Form 1-Bit Mask Construction}
\label{subsec:phase_quant_mask}
We now derive a closed-form ON/OFF mask that steers energy toward a desired
{target} AoD $\theta_T$ (or multiple targets) under incidence $\theta_I$.

\subsubsection{Ideal continuous phase profile}
To phase-align all active contributions toward $\theta_T$, an ideal programmable
reflector would apply the compensating phase
\begin{equation}
\phi_m \triangleq kx_m\big(\sin\theta_T+\sin\theta_I\big),
\label{eq:psi_ideal}
\end{equation}
so that the terms in \eqref{eq:pb_sum} add constructively at $\theta=\theta_T$.

\subsubsection{Cosine-threshold ON/OFF mapping}
\begin{figure}[t]
\centering
\includegraphics[width=0.75\columnwidth]{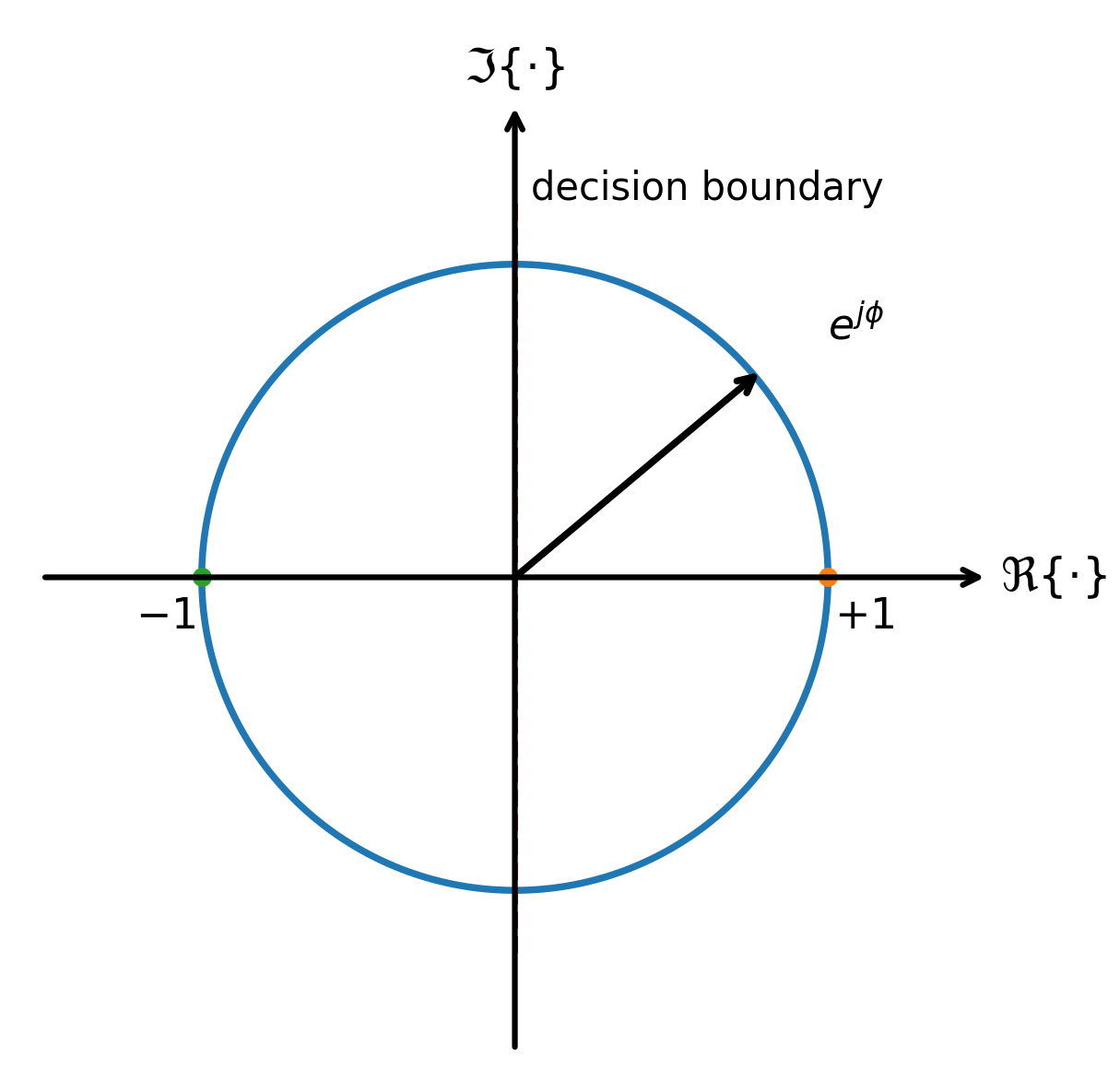}
\caption{Geometric view of 1-bit phase quantization on the unit circle. The ideal phasor $e^{j\phi}$ is mapped to $\{+1,-1\}$ using the sign of $\Re\{e^{j\phi}\}=\cos\phi$. The fabricated reflector implements the associated ON/OFF mask via metallization (ON) or suppression (OFF).}
\label{fig:unit_circle}
\end{figure}

We first quantize the ideal unit-modulus phasor $e^{j\phi_m}$ onto the bipolar
alphabet $\{+1,-1\}$ via nearest-neighbor projection as follows (see Fig. \ref{fig:unit_circle})
\begin{equation}
\begin{aligned}
\hat{w}_m
&=\arg\min_{w\in\{+1,-1\}}\left|e^{j\phi_m}-w\right|^2
=\mathrm{sgn}\!\big(\Re\{e^{j\phi_m}\}\big)\\
&=\mathrm{sgn}\!\big(\cos\phi_m\big)\in\{+1,-1\}.
\end{aligned}
\label{eq:w_quant}
\end{equation}
The bipolar alphabet corresponds to a 1-bit $0/\pi$ reflection-phase surface. 
Realizing $\hat{w}_m=-1$ in passive hardware requires a reflection phase shift of
approximately $\pi$ with comparable magnitude, which cannot be obtained by
metallization presence/absence alone. Achieving a true $-1$ state therefore
requires an additional passive modulation mechanism (e.g., geometry-encoded phase
such as thickness/height modulation over a ground plane), which is outside the
scope of the fabrication approach used here.

Since the fabricated surface in this work realizes a binary amplitude state, we map the
bipolar code to an ON/OFF mask by assigning $+1\mapsto 1$ (ON) and $-1\mapsto 0$ (OFF) as follows
\begin{equation}
b_m \triangleq \frac{\hat{w}_m+1}{2}
\qquad \Longleftrightarrow \qquad
b_m=\mathbf{1}\{\cos\phi_m\ge 0\},
\label{eq:b_from_w}
\end{equation}
{  {where $\mathbf{1}\{\cdot\}$ denotes the indicator function, i.e.,
$\mathbf{1}\{\mathcal{A}\}=1$ if the condition $\mathcal{A}$ is true and
$\mathbf{1}\{\mathcal{A}\}=0$ otherwise.}}
Thus, the mask retains locations whose ideal phases lie in the constructive
half-plane and suppresses those that would contribute destructively at $\theta_T$.

\subsubsection{Extension to multiple target directions}
For $L$ target AoDs $\{\theta_{T,\ell}\}_{\ell=1}^{L}$ with complex weights
$\{\alpha_\ell\}$, we superpose the ideal phase ramps and threshold the real part to obtain
\begin{align}
s_m &\triangleq \sum_{\ell=1}^{L}\alpha_\ell\,
\exp\!\big(jk(\sin\theta_{T,\ell}+\sin\theta_I)x_m\big),
\label{eq:Sm_multibeam}\\
b_m &\triangleq \mathbf{1}\{\Re\{s_m\}\ge 0\}.
\label{eq:bm_multibeam}
\end{align}
This produces a single binary pattern that can create multiple deterministic
non-specular lobes.

\subsubsection{Global phase offset for finite-aperture robustness}
For finite $M$, the thresholding in \eqref{eq:b_from_w} can be sensitive to how the
discrete samples fall relative to the decision boundary $\cos(\phi_m)=0$. We
therefore allow a constant phase offset $\psi$ prior to thresholding as follows
\begin{equation}
b_m(\psi)\triangleq \mathbf{1}\{\cos(\phi_m+\psi)\ge 0\}.
\end{equation}
The offset does not change the phase-ramp slope (hence does not change the
target geometry), but it can improve finite-aperture performance by shifting the
threshold relative to the sample phases. Unless otherwise stated, we set $\psi=0$ in this paper.

\subsection{Thinning Ratio (ON Fraction)}
\label{subsec:thinning_and_efficiency}
A direct consequence of ON/OFF masking is aperture thinning. Define the ON
fraction (thinning ratio)
\begin{equation}
\eta_M \triangleq \frac{1}{M}\sum_{m=0}^{M-1} b_m .
\label{eq:etaM_def}
\end{equation}

\begin{lemma}[Asymptotic $50\%$ thinning ratio for the single-target cosine-threshold mask]
\label{lem:TT}
Consider $b_m=\mathbf{1}\{\cos(\phi_m)\ge 0\}$ with phase progression
$\phi_{m+1}-\phi_m=k d_0(\sin\theta_T+\sin\theta_I)$. If
\begin{equation}
\frac{k d_0(\sin\theta_T+\sin\theta_I)}{\pi}\notin \mathbb{Q},
\label{eq:irrational_condition}
\end{equation}
then
\begin{equation}
\lim_{M\to\infty}\eta_M=\frac{1}{2}.
\label{eq:eta_half}
\end{equation}
\end{lemma}
\noindent\textit{Proof:} See Appendix~\ref{app:proof_lemma1}.

Lemma~\ref{lem:TT} shows that the cosine-threshold construction activates
asymptotically half of the available locations for generic steering geometries.
Condition~\eqref{eq:irrational_condition} excludes only degenerate cases where the
phase increment is exactly commensurate with the $\pi$-periodic sign structure of
the cosine threshold.  Fig.~\ref{fig:lemma2_illustration}(a) illustrates Lemma~\ref{lem:TT} by showing that the ON fraction (thinning ratio)
$\eta_M$ converges toward $0.5$ as the number of elements increases for a representative steering geometry.

\subsection{Distribution-Free Target-Lobe Gain Bound}
\label{sub_E}
We next quantify the achievable {target-direction} gain at $\theta=\theta_T$.
Define the coherent sum toward the target as
\begin{equation}
S_M(\mathbf b)\triangleq \sum_{m=0}^{M-1} b_m e^{-j\phi_m},\qquad
S_M^\star \triangleq \max_{\mathbf b\in\{0,1\}^M}\big|S_M(\mathbf b)\big|,
\label{eq:coherent_sum_onoff}
\end{equation}
and the normalized {target-lobe} gain as
\begin{equation}
\gamma \triangleq \frac{|S_M(\mathbf b)|^2}{M^2},\qquad
\gamma^\star \triangleq \max_{\mathbf b\in\{0,1\}^M}\gamma
= \frac{(S_M^\star)^2}{M^2}.
\label{eq:gamma_def}
\end{equation}
Lemma~\ref{lem:onoff_lb} establishes a distribution-free lower bound on the best achievable normalized target-lobe gain under the ON/OFF masking constraint.
\begin{lemma}[Distribution-free gain bound for single-target ON/OFF masking]
\label{lem:onoff_lb}
For arbitrary element locations $\{x_m\}$ and any incidence/target angles
$(\theta_I,\theta_T)$, the maximum achievable normalized target-lobe gain satisfies
\begin{equation}
\gamma^\star \ge \frac{1}{\pi^2}.
\label{eq:onoff_amp_lb} 
\end{equation}
\end{lemma}
\noindent\textit{Proof:} See Appendix~\ref{app:onoff_proof}.

\begin{figure}[t]
\centering
\includegraphics[width=0.9\linewidth]{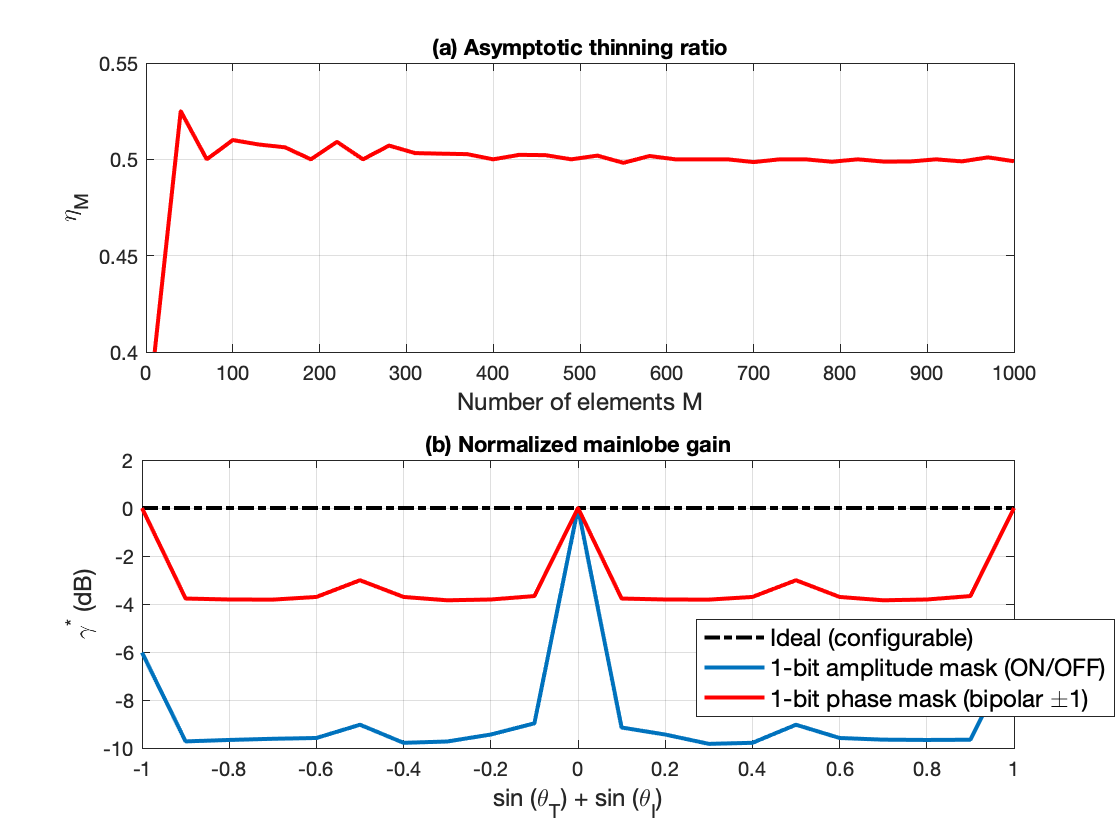}
\caption{(a) Thinning ratio $\eta_M$ versus number of elements for $\theta_I=60^\circ$ and $\theta_T=-30^\circ$,
illustrating convergence to $0.5$ (Lemma~\ref{lem:TT}). (b) Normalized {target-lobe} gain
$\gamma=|S_M|^2/M^2$ versus $\sin\theta_T+\sin\theta_I$ for $M=64$ and $d/\lambda=0.5$. ON/OFF masking incurs a
worst-case loss of $9.94$~dB relative to ideal continuous-phase control and $\approx 6$~dB relative to $\pm1$ control,
consistent with Lemma~\ref{lem:onoff_lb}.}
\label{fig:lemma2_illustration}
\end{figure}

Lemma~\ref{lem:onoff_lb} gives a worst-case guarantee on target-direction performance under the ON/OFF masking constraint. Specifically, the maximum normalized target-lobe gain is always at least $1/\pi^2$, which corresponds to a worst-case loss of $9.94$~dB relative to the ideal continuous-phase upper bound.  Fig.~\ref{fig:lemma2_illustration}(b) illustrates this result by comparing the normalized target-lobe gain of ON/OFF masking with bipolar ($\pm1$) and ideal continuous-phase control. As expected, $\pm1$ control incurs a smaller loss (about $3.9$~dB), since sign reversals can re-phase contributions that ON/OFF masking must suppress; a similar $3.9$~dB 1-bit quantization loss was reported in~\cite{WuZhang_TCOM2020_DiscreteRIS}, though this work uses a closed-form 1-bit assignment rule. The figure also confirms the $9.94$~dB worst-case ON/OFF loss established by Lemma~\ref{lem:onoff_lb}. Finally, at $\sin\theta_T+\sin\theta_I=0$, all schemes achieve $\gamma=1$ (0~dB),  corresponding to the specular-reflection condition.

\section{Diffraction-Order (Grating-Lobe) Beam Steering via Uniform Period Selection}
\label{sec:grating_optim}
Section~\ref{sec:problem_mask} introduced fixed-aperture ON/OFF masking on a dense
$d_0\approx \lambda/2$ scaffold. Here we present a complementary passive mechanism
based on uniform-period activation. The key design parameter is the inter-element
spacing (period) $\delta$ between adjacent active reflecting elements, with
$\delta\ge d_0$. In practice, $\delta$ can be realized by activating elements with
a periodic pattern on the dense lattice (e.g., activating every $m$th lattice
location along the synthesis dimension so that $\delta=\kappa d_0$). A uniformly
periodic aperture produces a discrete set of reflected lobes indexed by an integer
diffraction order $n$.  By selecting $(\delta,n)$, a chosen order (grating lobe) can be placed at a
desired target departure angle using closed-form relationships.

\subsection{Uniform-Period Array-Factor Model}
\label{gp}
Consider a linear reflector of $M$ identical reflecting elements with uniform
inter-element spacing $\delta$ (distance between adjacent active elements). Using
the centered indexing in \eqref{eq:xm_sys}, the element coordinates are
\begin{equation}
x_m \triangleq \left(\frac{M-1}{2}-m\right)\delta,\qquad m=0,1,\ldots,M-1.
\label{eq:xm_uniform}
\end{equation}
A plane wave impinges from AoA $\theta_I$ and the scattered field is observed
toward AoD $\theta_T$ in the reflection half-space.  Generalizing
\eqref{eq:p_sum} to uniform weights ($[\mathbf{\Psi}]_{m,m}=1$) gives
\begin{equation}
p(\theta_T,\theta_I)=\sum_{m=0}^{M-1} e^{-jk x_m(\sin\theta_T+\sin\theta_I)}.
\label{eq:p_uniform_sum}
\end{equation}

\subsection{Diffraction Orders and Single-Beam Steering}
For a uniformly periodic aperture, \eqref{eq:p_uniform_sum} admits the closed form \cite{trees}
\begin{equation}
p(\theta_T,\theta_I)=
e^{-j\frac{(M-1)}{2}\delta(k_I+k_T)}
\frac{\sin\!\left(\frac{M}{2}\delta(k_I+k_T)\right)}
{\sin\!\left(\frac{1}{2}\delta(k_I+k_T)\right)},
\label{eq:p_uniform_closed}
\end{equation}
where $k_I \triangleq k\sin\theta_I$ and $k_T \triangleq k\sin\theta_T$.
The principal maxima occur when the inter-element phase increment is an integer
multiple of $2\pi$,  i.e.  $
\frac{\delta}{\lambda}\big(\sin\theta_I+\sin\theta_T\big)=n,  n\in\mathbb{Z}.$
Each integer $n$ defines a diffraction order (a candidate reflected lobe). The
departure angle of order $n$ can be written as
\begin{equation}
\theta_n=\sin^{-1}\!\left(n\frac{\lambda}{\delta}-\sin\theta_I\right).
\label{eq:theta_n_uniform}
\end{equation}

\subsubsection{Closed-form period for steering to a target AoD}
To align order $n$ with a desired AoD $\theta_T$, enforce $\theta_n=\theta_T$ in
\eqref{eq:theta_n_uniform} to obtain 
\begin{equation}
\delta^{\ast}(n)=\frac{n\lambda}{\sin\theta_T+\sin\theta_I}.
\label{eq:delta_star_n}
\end{equation}
For single-beam steering,  we set $n=\pm 1$ to minimize the distance.  For $n=1$,
\begin{equation}
\delta^{\ast}=\frac{\lambda}{\sin\theta_T+\sin\theta_I}.
\label{eq:delta_star_1}
\end{equation}

\subsection{Multi-Beam Operation via Multiple Visible Diffraction Orders}
Multiple diffraction orders may be visible in the far field. Order $n$ is visible
only if the argument of $\sin^{-1}(\cdot)$ in \eqref{eq:theta_n_uniform} lies in
$[-1,1]$,  and hence 
\begin{equation}
\left|\,n\frac{\lambda}{\delta}-\sin\theta_I\,\right|\le 1.
\label{eq:visible_condition}
\end{equation}
The corresponding integer range of visible orders becomes
\begin{equation}
n_{\min}=\left\lceil \frac{\delta}{\lambda}\left(\sin\theta_I-1\right)\right\rceil,
\qquad
n_{\max}=\left\lfloor \frac{\delta}{\lambda}\left(\sin\theta_I+1\right)\right\rfloor,
\label{eq:n_bounds}
\end{equation}
with $N_{\text{orders}}=n_{\max}-n_{\min}+1$. If a second target AoD coincides with
one of the visible $\theta_n$, the same $\delta$ illuminates it; otherwise, $\delta$
and/or the chosen order can be redesigned subject to physical constraints.

\section{3D-Printed ``Inkwell'' Prototypes and Stencil-Based Fabrication}
\label{sec:fabrication}
This section describes the low-cost fabrication workflow used to realize and
experimentally validate the two fully passive beam-shaping mechanisms developed in this paper, namely
(i) fixed-aperture 1-bit spatial masking (Section~\ref{sec:problem_mask}) and
(ii) diffraction-order (uniform-period) steering (Section~\ref{sec:grating_optim}).
Both designs are implemented on a common 3D-printed ``inkwell'' reflector platform, in which a
planar aperture is discretized into a dense lattice of candidate metallization locations.    {The reflector realization involves three components. The first is the 3D-printed inkwell scaffold,
which serves as the physical base platform. The second is the analytically derived binary mask
pattern, which specifies the ON/OFF metallization layout. The third is the stencil, which is used
during fabrication to deposit conductive ink only into the selected wells. Since the beam synthesis
in this paper is carried out for azimuthal steering with the elevation angle fixed to zero, the
fabricated prototypes are realized on a 2-D lattice by replicating the same 1-D azimuthal pattern
across rows. The reflector dimensions, lattice parameters, material choices, and operating
frequency used in this paper are summarized in Table}~\ref{tab:fabrication_specs}.

\begin{figure}[t]
\hspace{-3mm}
\includegraphics[width=1.03\linewidth]{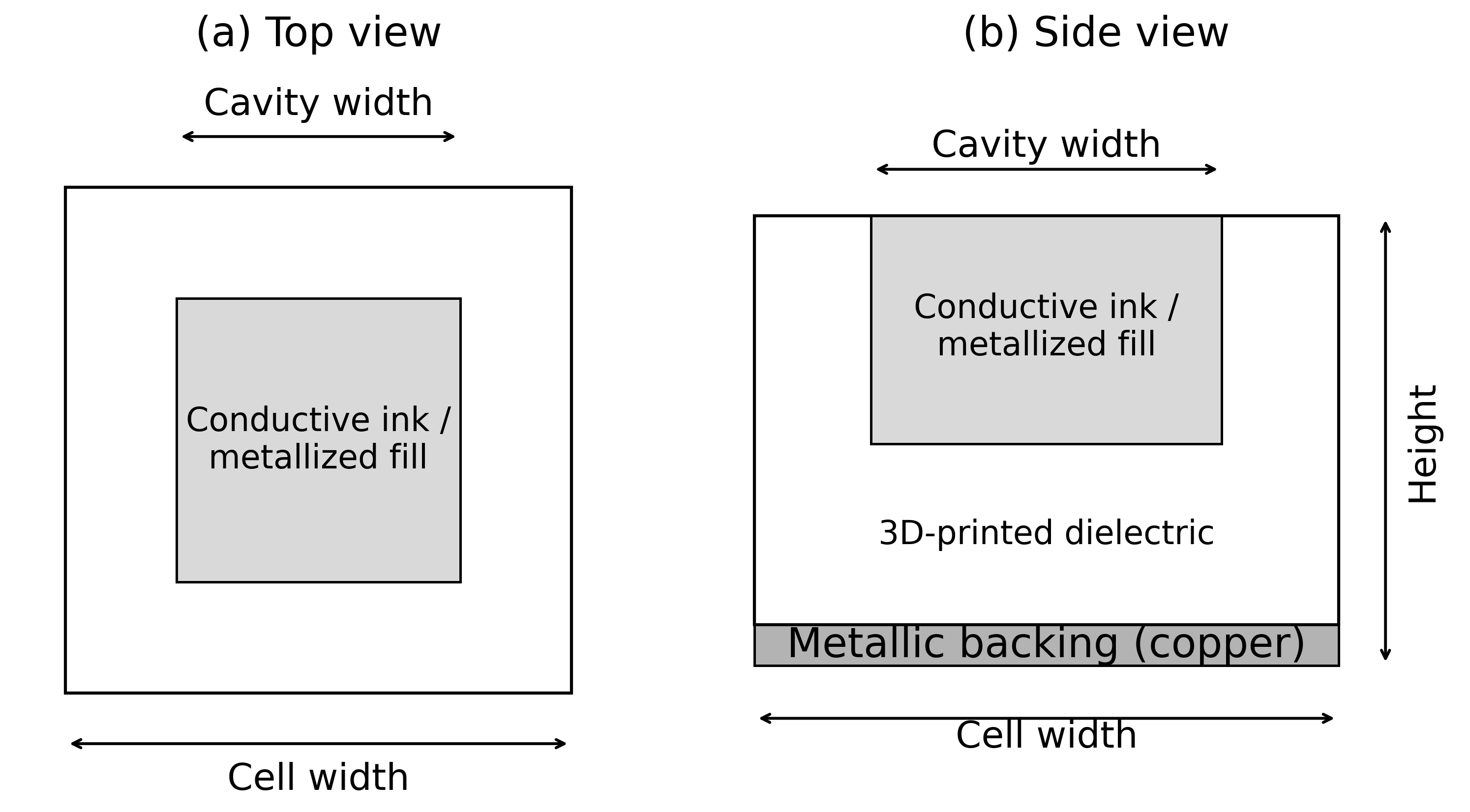}
\caption{  {Representative reflector cell (element) geometry: (a) top view and (b) side view. The cell consists of a conductive ink/metallized fill formed inside the inkwell cavity, a 3D-printed dielectric scaffold, and a metallic copper backing in direct contact with the dielectric.  }}
\label{fig:view}
\end{figure}

Fig.~\ref{fig:view}   {shows a representative reflector cell in top and side view. Each cell consists
of a conductive ink/metallized fill inside an inkwell cavity, a 3D-printed dielectric scaffold, and
a metallic copper backing realized using copper adhesive tape directly attached to the dielectric.
This yields a metal-backed dielectric structure similar in structural spirit to prior passive
reflector implementations such as MilliMirror} \cite{qian2022millimirror},    {although the present work
does not rely on thickness-tuned unit-cell phase synthesis.}

  {The fabricated reflectors are designed using the aperture-level analytical array model developed in
this paper. In this model, each lattice location is treated as an effective reflector sample
contributing to the reradiated field. The active (metallized) locations are selected either by the
cosine-threshold rule described in Section}~\ref{subsec:phase_quant_mask} or by the periodic-spacing
selection described in Section~\ref{gp}.   {These analytical results are used both to design the
fabricated reflectors and to generate the corresponding theoretical patterns. Accordingly, this work
does not include full-wave extraction of element-level reflection magnitude and phase under normal or
oblique incidence,  as would be done in a Computer Simulation Technology (CST) or High Frequency Structure Simulator (HFSS)-based metasurface unit-cell analysis.  Similar
element-/surface-level abstractions are also common in RIS system modeling }
\cite{khateeb_LiS,HuangRIS2019}.    {In the analytical model, metallized cells are treated as ideal
reflecting points, while non-metallized cells are assumed to contribute negligibly to the reflected
field. Full-wave element-level reflection characterization under normal and oblique incidence is left
for future work.}

\subsection{Common ``Inkwell'' Geometry and Relation to the Array Model}
\begin{figure}[t]
\centering
\subfloat[\footnotesize Bare inkwell base.]{
  \includegraphics[width=0.28\linewidth]{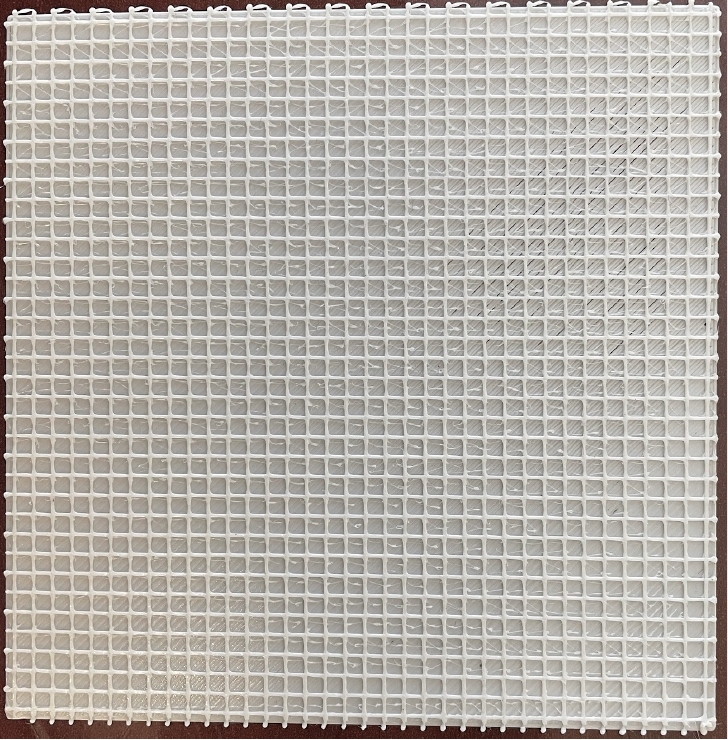}
  \label{fig:fullgrid_bare}
}
\hfill
\subfloat[\footnotesize All-ON front.]{
  \includegraphics[width=0.29\linewidth]{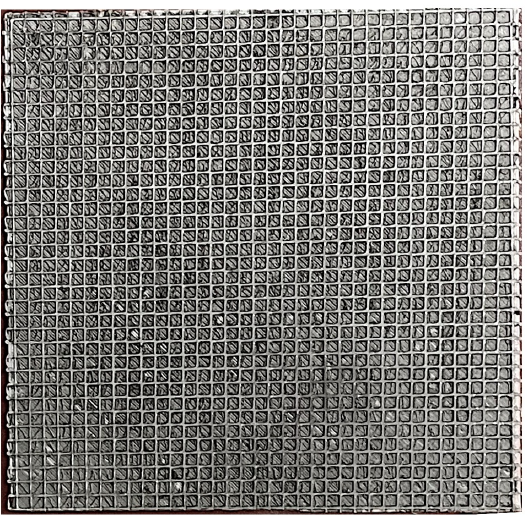}
  \label{fig:fullgrid_front}
}
\hfill
\subfloat[\footnotesize All-ON back.]{
  \includegraphics[width=0.29\linewidth]{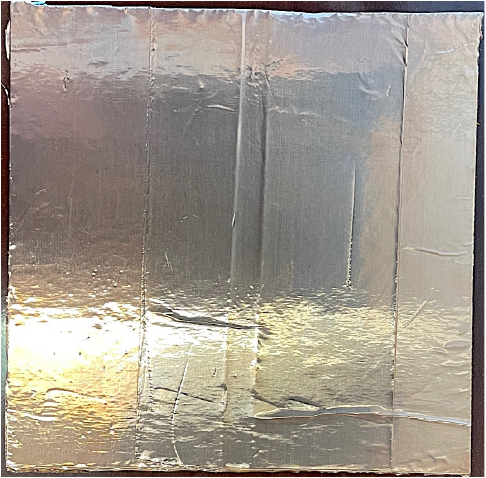}
  \label{fig:fullgrid_back}
}
\caption{All-ON reference reflector: (a) bare base, (b) metallized front, (c) copper-backed rear.}
\label{fig:full_grid_reflector}
\end{figure}

Fig.~\ref{fig:full_grid_reflector}(a) shows the bare 3D-printed inkwell base used in this paper. The aperture is discretized into a
uniform candidate lattice with pitch $d_0=2.5~\mathrm{mm}$. Each candidate location is a square inkwell with a
$2.2\times 2.2~\mathrm{mm}$ opening and depth $\approx 0.4~\mathrm{mm}$, printed into a dielectric base of total
thickness $\approx 0.8~\mathrm{mm}$ (approximately $0.4$~mm bottom floor plus $0.4$~mm wall layer). The well walls
(i) confine conductive ink during deposition and (ii) enforce a repeatable metallized footprint across prototypes.

From a modeling perspective, the inkwell grid provides a fixed sampling scaffold for implementing the diagonal
aperture interaction $\mathbf{\Psi}$ in the array-factor model. For spatial masking, a subset of wells is metallized
to realize $b_m\in\{0,1\}$, thereby implementing the binary aperture described in
Section~\ref{sec:problem_mask}. For diffraction-order steering, entire columns  are metallized with a
uniform effective spacing $\delta$ to implement the periodic aperture described in
Section~\ref{sec:grating_optim}. Metallization is implemented by filling selected inkwells with a commercial
water-based conductive paint~\cite{conductive_paint}. Fig.~\ref{fig:full_grid_reflector}(b) shows the resulting
all-ON configuration, where all wells are metallized.  Since this paper targets azimuthal beam shaping, the
designed 1-D pattern along the $x$-axis is replicated across the $y$-axis (constant vertical pitch), producing a
striped 2-D aperture.

\begin{table}[t]
\centering
\caption{Geometry, material, and fabrication parameters of the 3D-printed ``inkwell'' reflector platform used in this paper.}
\label{tab:fabrication_specs}
\renewcommand{\arraystretch}{1.08}
\begin{tabular}{p{0.33\columnwidth} p{0.49\columnwidth}}
\hline
\textbf{Parameter} & \textbf{Value / Description} \\
\hline
Aperture size & $90 \times 90$~mm (square), centered at $(0,0)$ \\
Candidate grid (scaffold) & $35 \times 35$ wells ($1225$ locations) \\
Center-to-center pitch & $d_0 = 2.5$~mm in $x$ and $y$ ($\approx \lambda/2$ at $60.48$~GHz) \\
Inkwell opening (front) & $2.2 \times 2.2$~mm square \\
Inkwell depth & $\approx 0.4$~mm (ink reservoir) \\
Base thickness & $\approx 0.8$~mm total \\
 & ($\sim$0.4~mm bottom dielectric floor + $\sim$0.4~mm wall layer) \\
Stencil thickness & $\approx 0.8$~mm \\
Stencil aperture & $2.1 \times 2.1$~mm square (slightly smaller than wells) \\
Substrate material & 3D-printed dielectric (PLA/PETG) \\
Metallization & SilexCore conductive paint~\cite{conductive_paint} deposited into wells (stencil-guided for masks) \\
Ground plane & Continuous copper tape/sheet on rear face (high reflectivity at 60.48~GHz) \\
\hline
\end{tabular}
\end{table}

\subsection{Copper-Foil Backing as a Passive Ground Plane}
All prototypes use a continuous copper-foil tape layer on the rear side of the printed base to form a conductive
ground plane (Fig.~\ref{fig:full_grid_reflector}(c)). This backing is critical at $60$~GHz as it provides a
high-conductivity termination that increases reflection efficiency and ensures that the front-side pattern
(modulated by ON/OFF metallization or periodic activation) perturbs a strong reflected field.

\subsection{Stencil Layer for Repeatable Binary Metallization}
\begin{figure}[t]
 \centering
 \includegraphics[width=0.45\textwidth]{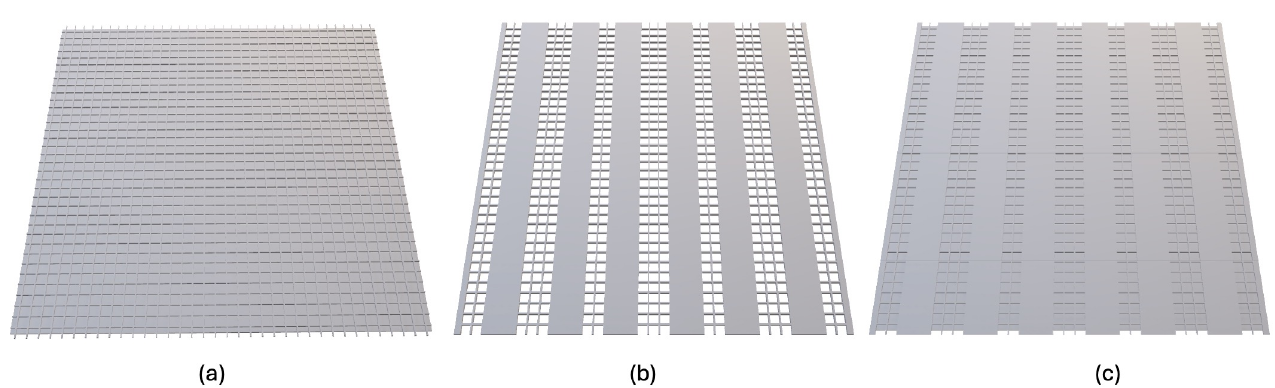}
 \caption{Inkwell base and matching stencil (CAD/STL views). (a) Base with all wells present. (b) Stencil used to metallize
 selected wells (ON locations). (c) Optional ``matched-base'' variant in which OFF locations are removed in the print;
 in this work, we use the base matched to the stencil (c)+ stencil(b),  but either (a)+(b) or (c)+(b) can be used.}
 \label{fig:stl3060}
\end{figure}

For spatial-masking prototypes, we fabricate a matching 3D-printed stencil that enforces the designed ON/OFF pattern
during conductive-ink deposition. The stencil (thickness $\approx 0.8~\mathrm{mm}$) contains square openings of
$2.1\times 2.1~\mathrm{mm}$ (slightly smaller than the $2.2\times 2.2~\mathrm{mm}$ well openings) to reduce edge
bleeding and improve pattern fidelity. The stencil is closed everywhere except at intended ON locations so that the
designed binary sequence is mapped into a repeatable metallization pattern. Fig.~\ref{fig:stl3060} shows representative   {computer-aided design (CAD) and stereolithography (STL) renderings} of the inkwell base and a corresponding stencil for a $(\theta_I,\theta_T)=(30^\circ,-60^\circ)$
design case.

\subsection{Prototype Classes and Fabrication Workflow}
Using the common printing-and-deposition workflow, we fabricate three prototype classes which correspond to the
designs analyzed in Sections~\ref{sec:problem_mask} and~\ref{sec:grating_optim}. Each prototype is realized by
(i) 3D-printing the inkwell base (and a matching stencil),
(ii) applying a continuous copper ground plane to the rear face, and
(iii) depositing conductive paint into the selected inkwells. For mask-coded designs, the stencil confines
deposition to the intended ON locations to improve pattern fidelity and repeatability.

The all-ON reference reflector in Fig.~\ref{fig:full_grid_reflector}(b) metallizes all $35\times 35$ wells and serves as a
baseline specular/reference response for measurement validation. For fixed-aperture 1-bit spatial masking, the distance
remains fixed at $d_0=2.5~\mathrm{mm}$ and beam control is achieved by metallizing a subset of wells according to the designed
binary mask. For diffraction-order steering, beam (lobe) placement is achieved by metallizing columns with a uniform
effective spacing $\delta$, with each active column replicated across rows to form a periodic 2-D pattern.

\begin{figure}[t]
\centering
\begin{minipage}{0.35\linewidth}
  \centering
  \includegraphics[width=\linewidth]{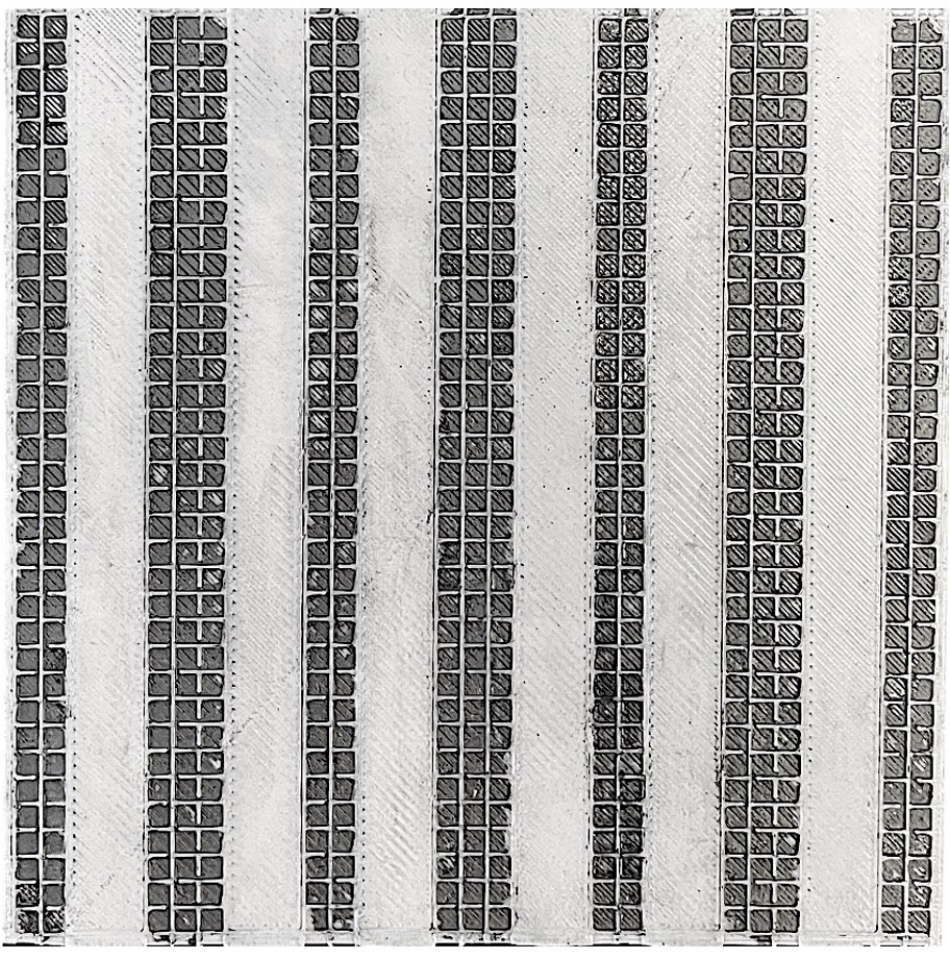}\\
  \scriptsize (a) ON/OFF, $(\theta_I,\theta_T)=(30^\circ,-60^\circ)$
\end{minipage}
\hspace{3mm}
\begin{minipage}{0.34\linewidth}
  \centering
  \includegraphics[width=\linewidth]{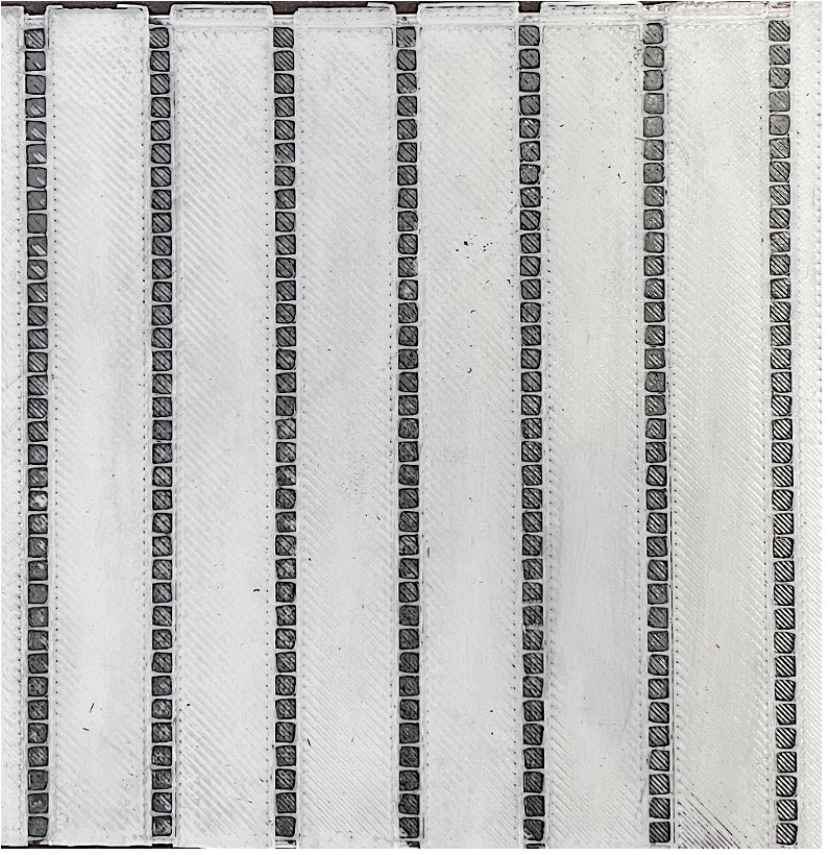}\\
  \scriptsize (b) Diffraction-order, $(\theta_I,\theta_T)=(30^\circ,-60^\circ)$
\end{minipage}

\vspace{1.5mm}

\begin{minipage}{0.35\linewidth}
  \centering
  \includegraphics[width=\linewidth]{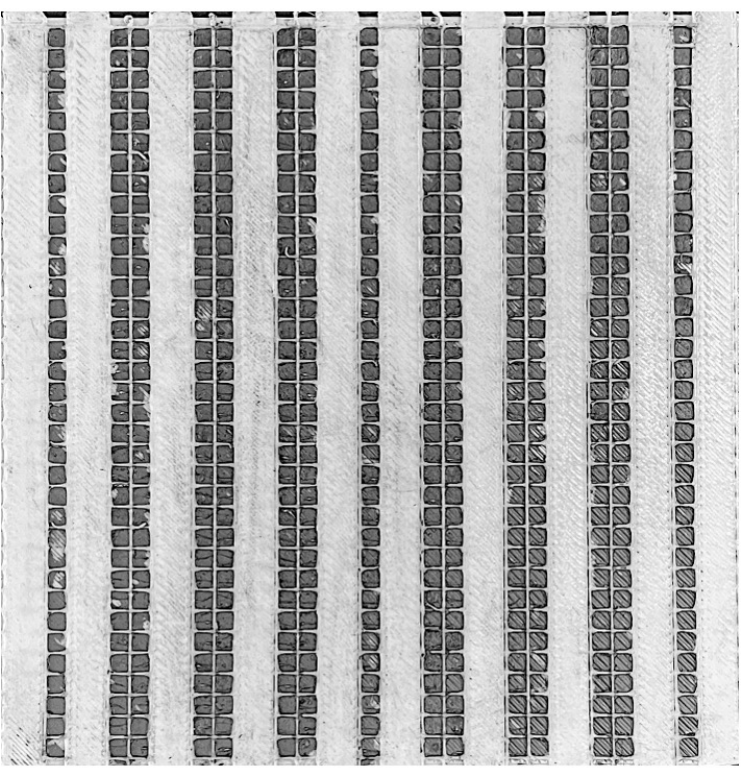}\\
  \scriptsize (c) ON/OFF, $(\theta_I,\theta_T)=(45^\circ,-10^\circ)$
\end{minipage}
\hspace{3mm}
\begin{minipage}{0.35\linewidth}
  \centering
  \includegraphics[width=\linewidth]{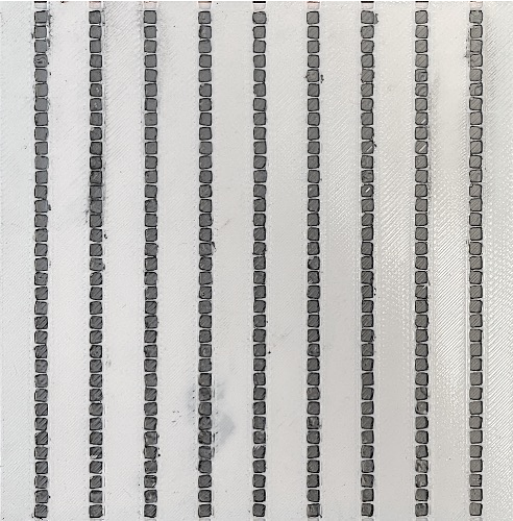}\\
  \scriptsize (d) Diffraction-order, $(\theta_I,\theta_T)=(45^\circ,-10^\circ)$
\end{minipage}
\caption{Fabricated prototypes on the inkwell substrate (ON/OFF mask vs.\ diffraction-order).}
\label{fig:combined_prototypes}
\end{figure}

Fig.~\ref{fig:combined_prototypes} shows fabricated prototypes for two representative steering configurations,
$(\theta_I,\theta_T)=(30^\circ,-60^\circ)$ and $(\theta_I,\theta_T)=(45^\circ,-10^\circ)$, comparing spatial-mask and
diffraction-order realizations on the same inkwell platform. These prototypes are used in Section \ref{sec:results} to validate
the proposed beam-control methods through $60.48$~GHz azimuthal sweeps.

\section{Performance Criteria and Evaluation Metrics}
\label{sec:metrics}

This section defines the metrics used to evaluate the fully passive, fabrication-coded
reflectors. The proposed apertures are
designed to synthesize deterministic {non-specular} departure lobes toward intended
user directions. 
We quantify (i) target-direction power delivery and (ii) system-level impact
under reflector-side operational-power constraints.

\section{Performance Criteria and Evaluation Metrics}
\label{sec:metrics}

This section defines the metrics used to evaluate the fully passive, fabrication-coded
reflectors developed in this paper. The proposed apertures are designed to generate
deterministic non-specular departure lobes toward intended user directions. Their
performance is quantified in terms of (i) target-direction response and
(ii) system-level impact under reflector-side operational-power constraints.

\subsection{Angular Pattern Metrics}
\label{subsec:target_metrics}
For a fixed incidence angle $\theta_I$, the theoretical angular power pattern is defined as
\begin{equation}
P(\theta;\theta_I)\triangleq \big|p(\theta,\theta_I)\big|^2,
\end{equation}
where $\theta$ denotes the departure angle in the reflection half-space. For the 1-D model
with $M$ elements, we define the normalized gain as follows
\begin{equation}
\gamma(\theta;\theta_I)\triangleq \frac{P(\theta;\theta_I)}{M^2},
\label{eq:gamma_metric}
\end{equation}
and for a desired target departure angle $\theta_T$, the target-direction power is
\begin{equation}
G_T \triangleq P(\theta_T;\theta_I).
\label{eq:GT_metric}
\end{equation}

\subsection{System-Level Mapping and Energy Efficiency}
\label{subsec:system_energy_metrics}
To connect target-direction performance to a system-level metric, we map the target-direction signal-to-noise ratio (SNR) to an achievable
rate using the Shannon formula
\begin{equation}
R(\theta_T)\triangleq B\log_2\!\big(1+\mathrm{SNR}(\theta_T)\big),
\label{eq:rate_metric}
\end{equation}
where $B$ is the communication bandwidth. In the theoretical comparisons, $\mathrm{SNR}(\theta_T)$ can be interpreted as
scaling with the target-direction gain $G_T$ in \eqref{eq:GT_metric}.  Energy efficiency is then computed as \cite{WangTCOM2024Power,WangEUSIPCO2023StaticPower,RihanNetwork2024_PassiveActiveRIS_ISAC}
\begin{equation}
\mathrm{EE}\triangleq \frac{R(\theta_T)}{P_{\mathrm{TX}}+P_{\mathrm{circ}}+P_{\mathrm{surf}}},
\label{eq:EE_metric}
\end{equation}
where $P_{\mathrm{TX}}$ is the radiated transmit power, $P_{\mathrm{circ}}$ is the fixed radio/baseband power excluding
$P_{\mathrm{TX}}$, and $P_{\mathrm{surf}}$ is the surface-side operational power required by the reflecting structure.
{Energy efficiency is a key deployment metric because it captures delivered throughput per unit total power and
explicitly accounts for surface-side overhead that impacts long-term operation in static
installations.} Combining \eqref{eq:rate_metric} and \eqref{eq:EE_metric} yields
\[
\mathrm{EE}(\theta_T)=\frac{B\log_2\!\big(1+\mathrm{SNR}(\theta_T)\big)}{P_{\mathrm{TX}}+P_{\mathrm{circ}}+P_{\mathrm{surf}}},
\]
which shows that $\mathrm{EE}$ increases with target-direction SNR through $R(\theta_T)$ and decreases with total power
consumption, particularly the surface-side overhead $P_{\mathrm{surf}}$ in powered RIS architectures. For the proposed
passive desings, no biasing or control circuitry is required after installation, and thus
$P_{\mathrm{surf}}\approx 0$ during operation.

\section{Results and Discussion} \label{sec:results} 
This section presents both theoretical and experimental results for the proposed fully passive coded reflectors. Section~\ref{subsec:theory_results} analyzes the performance of 1-bit ON/OFF masking and diffraction-order steering using the proposed aperture-level models. Section~\ref{subsec:exp_results} then reports over the air measurements and discusses their agreement with theory, along with implications for practical deployment. 

The 1-D array-factor expression in \eqref{eq:p_sum} is used to generate the azimuth-plane
plots, whereas the planar-aperture model in \eqref{eq:p_2d} is used to generate the
corresponding 2-D angular maps and 3-D patterns. For the ON/OFF design, the ideal
target-direction phase profile is first formed using \eqref{eq:psi_ideal}, then converted
to a binary mask using the cosine-threshold rule in \eqref{eq:b_from_w}, and finally
evaluated in the array model. The resulting binary vector $\mathbf{b}$ directly determines
the metallized cell locations in the fabricated reflector, i.e., entries with $b_m=1$
identify the wells that are filled with conductive ink, while entries with $b_m=0$ remain
unmetallized. Thus, the same analytically derived mask is used both to generate the
theoretical ON/OFF plots and to define the metallization pattern of the 3D-printed
prototype.

For the diffraction-order design, the required period is first obtained from the closed-form
design rule in \eqref{eq:delta_star_1}, and the resulting angular response is then computed
using the uniform-period model in \eqref{eq:p_uniform_sum}. That analytically determined
period is also used to set the spacing of the active metallized columns in the fabricated
3D-printed diffraction-order prototype. In the 2-D evaluations, the same 1-D azimuthal
design is replicated across rows to form the corresponding 2-D coded aperture.
Accordingly, all theoretical plots reported in this section are analytical evaluations of
the proposed aperture-level models, and the same analytical design parameters are used to
define the physical metallization pattern and spacing of the fabricated prototypes.

{  { To position the proposed approach relative to closely related designs, several benchmark cases are used throughout the theoretical and experimental evaluation. In the theoretical results, the proposed ON/OFF reflector is compared against three reference designs: a 1-bit ($\pm1$) benchmark motivated by discrete-state RIS literature}} \cite{WuZhang_TCOM2020_DiscreteRIS},   { a conventional all-ON passive reflector baseline representing non-engineered specular reflection as in prior passive mmWave reflector studies} \cite{khawaja2020coverage},   {and an ideal continuous-phase reflector serving as an upper-bound reference for target-direction gain, consistent with common RIS/LIS benchmarking practice} \cite{khateeb_LiS,HuangRIS2019}.   {In the experimental results, the fabricated designs are compared against both the passive all-ON reflector and a smooth copper plate of the same aperture size. The copper plate serves as a conductor reference to verify that the all-ON metallized inkwell reflector exhibits comparable specular behavior. These benchmarks help clarify that the contribution of this work is a fully passive, fabrication-friendly coded-aperture reflector framework that enables target-direction beam shaping without requiring active tuning circuitry or a new 3-D geometry redesign for each beam objective.}
\subsection{Theoretical Results}
\label{subsec:theory_results}

Unless otherwise stated, all theoretical results and figures in this section are obtained
by evaluating the proposed analytical aperture-level models in MATLAB. No full-wave
unit-cell simulations are used in this work. The ON/OFF results
are generated using the formulation in Section~\ref{sec:problem_mask}, whereas the
diffraction-order results are generated using the periodic steering formulation in
Section~\ref{sec:grating_optim}. Unless otherwise stated, the analysis is carried out at
60~GHz using $M=35$ and $d_0/\lambda=0.5$.

\subsubsection{Single-target synthesis: $\theta_I=45^\circ$, $\theta_T=-10^\circ$}
\begin{figure}[t]
\centering
\includegraphics[width=1\columnwidth]{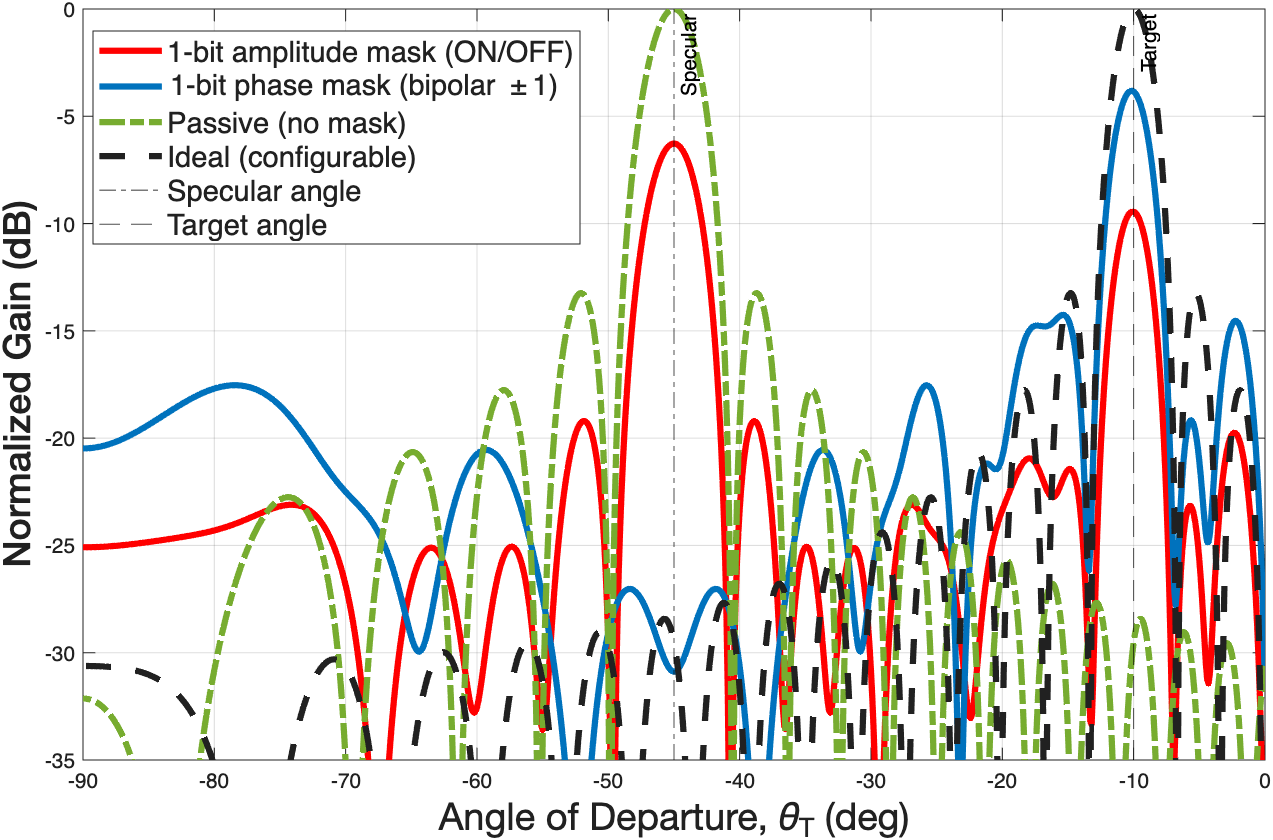}
\caption{Theory (60\,GHz, $\theta_I=45^\circ$, target $\theta_T=-10^\circ$, $M=35$, $d_0/\lambda=0.5$). Normalized gain $\gamma(\theta;\theta_I)$ for all-ON, 1-bit ON/OFF (cosine-threshold), 1-bit bipolar ($\pm1$) (cosine-threshold), and an ideal continuous-phase reflector.}
\label{fig:4510_32_fig3}
\end{figure}

Fig.~\ref{fig:4510_32_fig3} compares the normalized gain $\gamma(\theta;\theta_I)$ for four reflector
configurations: the conventional all-ON passive aperture, the proposed ON/OFF mask, a bipolar ($\pm1$)
1-bit reflector included as a discrete-state benchmark \cite{WuZhang_TCOM2020_DiscreteRIS}, and an ideal
continuous-phase reflector included as an upper-bound reference \cite{khateeb_LiS,HuangRIS2019}. The
all-ON aperture exhibits the expected specular-dominated response, with its strongest lobe at
$\theta_{\mathrm{spec}}=-\theta_I=-45^\circ$ and only weak non-specular reflection.  By contrast, the
proposed ON/OFF mask produces a deterministic non-specular lobe toward $\theta_T=-10^\circ$ by activating
a subset of lattice locations according to the cosine-threshold rule.

The non-specular target response of ON/OFF masking exhibits a $ 9.94$~dB loss relative to the ideal continuous-phase
upper bound. This is consistent with Lemma~2, which lower-bounds the maximum achievable normalized target-lobe gain by $1/\pi^2$
(i.e., $10\log_{10}(1/\pi^2)= -9.94$~dB). This loss is fundamentally due to amplitude-only control where an ideal surface
can phase-align all $M$ contributions at the target.   ON/OFF masking can only retain or suppress elements and cannot
re-phase unfavorable contributions. At the specular direction, the ON/OFF pattern incurs a smaller loss (approximately
$6.9$~dB) because the all-ON aperture is already phase-aligned and applying a non-specular mask primarily reduces the effective
aperture through thinning and redistributes energy into other angles. In particular, Lemma~1 shows an asymptotic
activation ratio near $50\%$, contributing an order-of-$3$~dB reduction, with the remaining loss attributable to the
non-uniform spatial weighting introduced to form the non-specular lobe.  The bipolar ($\pm1$) mask allows sign reversals and therefore more closely approximates the ideal coherent sum at the
target, yielding an $\approx 3.9$~dB loss, consistent with 1-bit phase
quantization trends reported in~\cite{WuZhang_TCOM2020_DiscreteRIS}.

\subsubsection{Diffraction-order steering (unconstrained aperture)}
\begin{figure}[t]
\centering
\includegraphics[width=1\columnwidth]{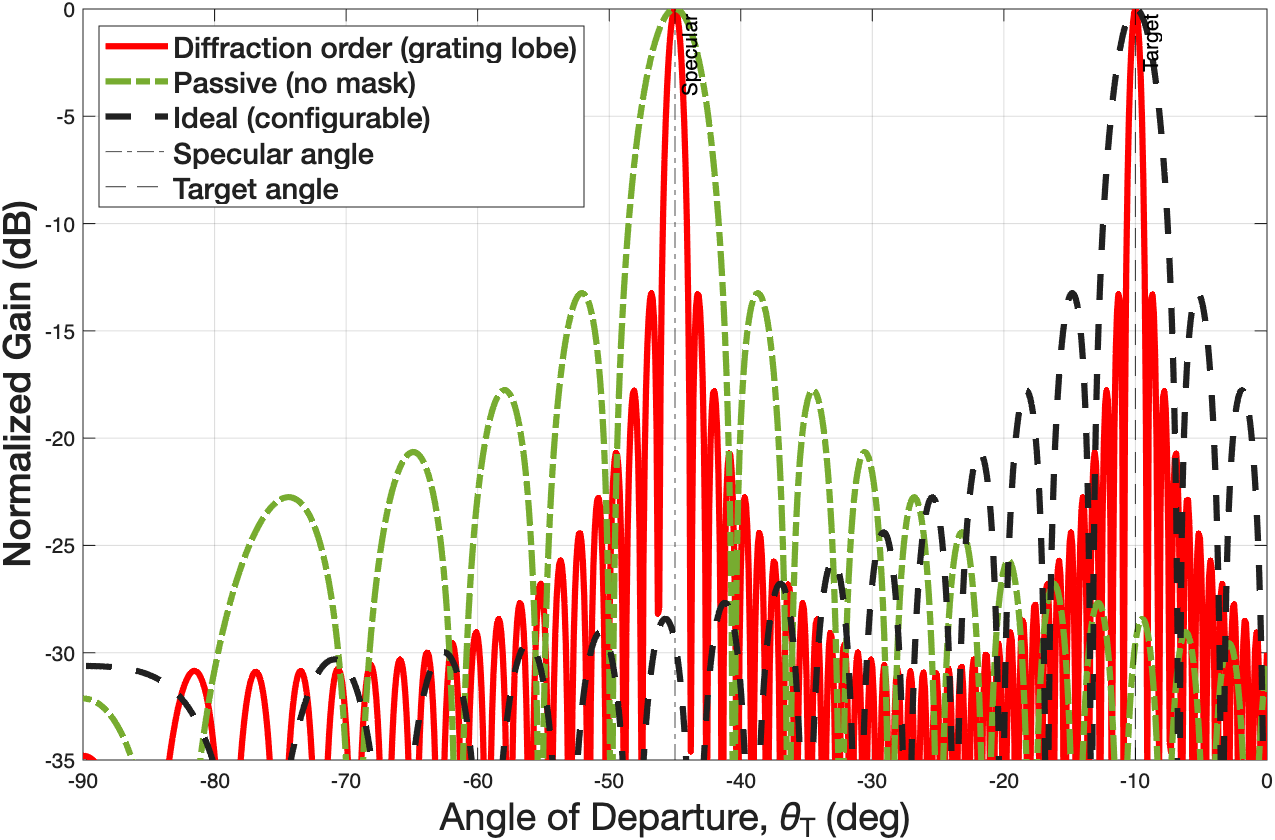}
\caption{Theory (60\,GHz, $\theta_I=45^\circ$, target $\theta_T=-10^\circ$, $M=35$). Diffraction-order steering with period $\delta=9.37$\,mm (all-ON), producing a grating lobe at $\theta_T$ while retaining the zeroth-order (specular) lobe.}
\label{fig:4510_32_fig5}
\end{figure}

Diffraction-order steering forms a non-specular lobe by enforcing a uniform effective period $\delta$ so that a chosen
diffraction order satisfies the grating condition for the desired departure angle $\theta_T$ (Section~\ref{sec:grating_optim}).
Fig.~\ref{fig:4510_32_fig5} illustrates $(\theta_I,\theta_T)=(45^\circ,-10^\circ)$ at 60\,GHz with $\delta=9.37$~mm, where
all $M=35$ elements are active on the periodic lattice. With this choice of $\delta$, a strong diffraction order is placed at
$\theta_T=-10^\circ$ while the zeroth-order (specular) response at $\theta_{\mathrm{spec}}=-45^\circ$ is retained, as expected
for a uniform-amplitude periodic aperture.

Because all elements participate coherently for the selected order, the target-lobe peak can approach the ideal
continuous-phase upper bound in the unconstrained-aperture setting shown in Fig.~\ref{fig:4510_32_fig5}. This gain advantage,
however, is achieved by increasing the inter-element spacing from the dense scaffold spacing ($d_0=2.5$~mm $\approx \lambda/2$)
to $\delta=9.37$~mm, which expands the physical aperture when $M$ is held fixed.  The increased aperture
length leads to narrower angular features (sharper main lobes and nulls) which can make alignment more sensitive in practice. 

\subsubsection{Multi-target synthesis: $\theta_I=30^\circ$, $\theta_T\in\{-7.8^\circ,-60^\circ\}$}
\begin{figure}[t]
\centering
\includegraphics[width=1\columnwidth]{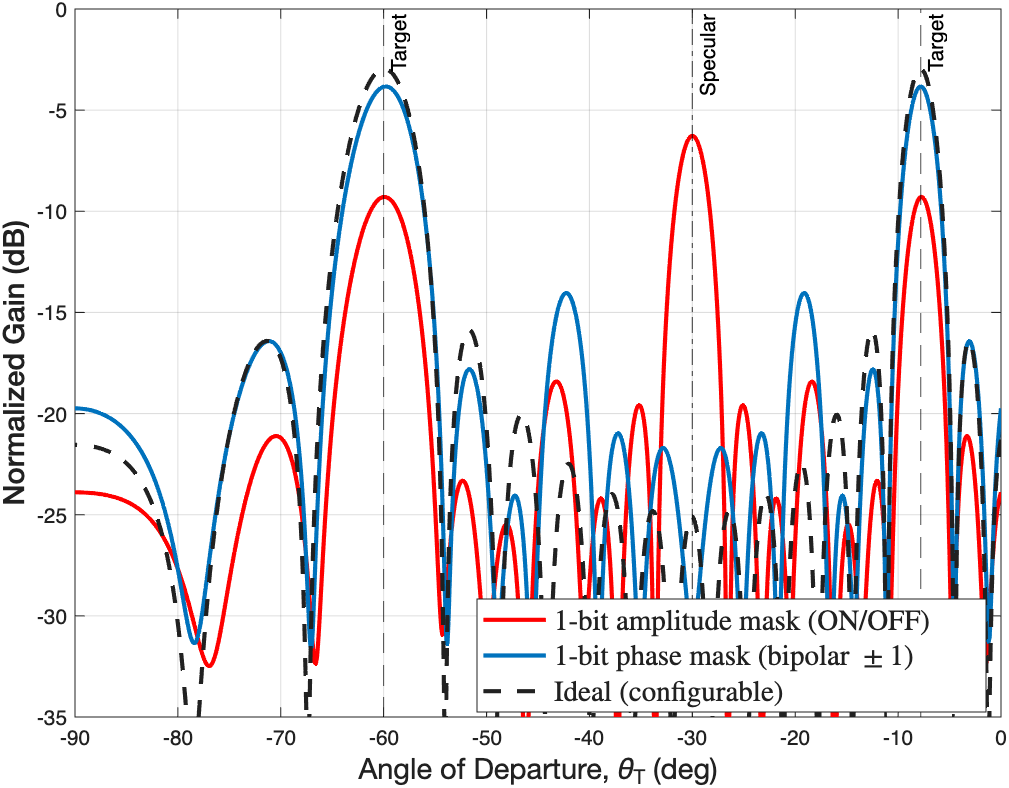}
\caption{Theory (60\,GHz, $\theta_I=30^\circ$, $M=35$): multi-beam synthesis targeting $\theta_T=-7.8^\circ$ and $-60^\circ$. Normalized gain $\gamma(\theta;\theta_I)$ is shown for the ON/OFF mask, the bipolar ($\pm1$) mask, and the ideal continuous-phase reflector.}
\label{fig:60830fig3}
\end{figure}

Fig.~\ref{fig:60830fig3} considers a two-target design obtained by complex superposition of the desired phase ramps followed
by binary thresholding as in \eqref{eq:Sm_multibeam}--\eqref{eq:bm_multibeam}. The ON/OFF mask produces two deterministic
non-specular lobes at $\theta_T=-7.8^\circ$ and $-60^\circ$ while retaining a residual specular component at
$\theta_{\mathrm{spec}}=-30^\circ$, which is expected under amplitude-only control. The bipolar ($\pm1$) mask, by allowing sign
reversals, more closely tracks the ideal continuous-phase reference and yields higher per-target coherence.

In the ideal continuous-phase case, the reflected energy is intentionally split across two directions, and therefore the
peak gain at each target is reduced relative to the single-target setting. The ON/OFF and bipolar curves exhibit the same
trend, with per-target peaks reduced by power sharing and by the 1-bit constraint. Notably, the characteristic loss behavior
observed in the single-target case is preserved.  When compared to the ideal continuous-phase reference,  the ON/OFF masking exhibits $\approx 10~\mathrm{dB}$ non-specular penalty
relative to ideal continuous-phase control,  while the
specular component experiences a lower reduction due to thinning and energy redistribution (as discussed for
Fig.~\ref{fig:4510_32_fig3}).

\begin{figure}[t]
\centering
\includegraphics[width=0.9\columnwidth]{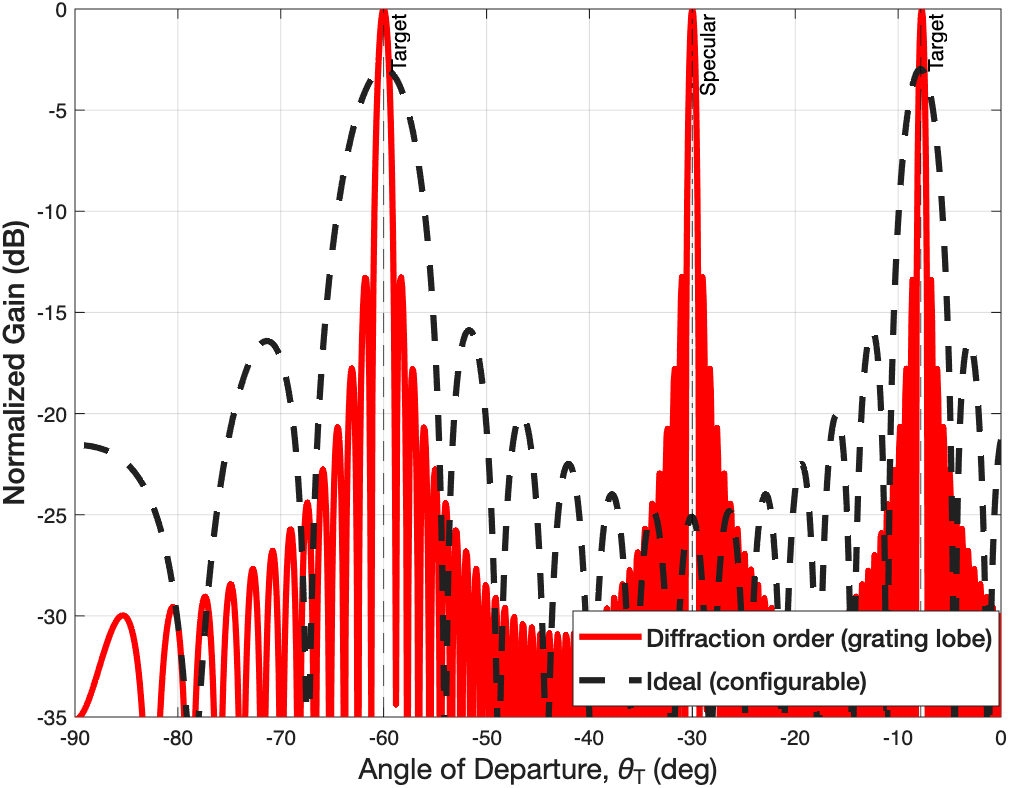}
\caption{Theory (60\,GHz, $\theta_I=30^\circ$, $M=35$): diffraction-order multi-lobe example with $\delta=13.66$\,mm, producing additional visible orders alongside the zeroth-order (specular) response.}
\label{fig:60830fig5}
\end{figure}

Fig.~\ref{fig:60830fig5} shows a diffraction-order multi-lobe example where a larger
effective period ($\delta=13.66$~mm) produces multiple visible diffraction orders alongside the zeroth-order (specular) lobe.
With $M$ held fixed, increasing $\delta$ expands the physical aperture and yields narrower angular profiles which can raise
the peak levels at certain directions.  This apparent gain improvement relative to ideal continuous-phase control ,  however, is primarily
a consequence of the increased aperture size rather than an intrinsic advantage of periodic steering under fixed-footprint
constraints.

\subsubsection{Aperture-matched comparison}
\begin{figure}[t]
\centering
\includegraphics[width=\columnwidth]{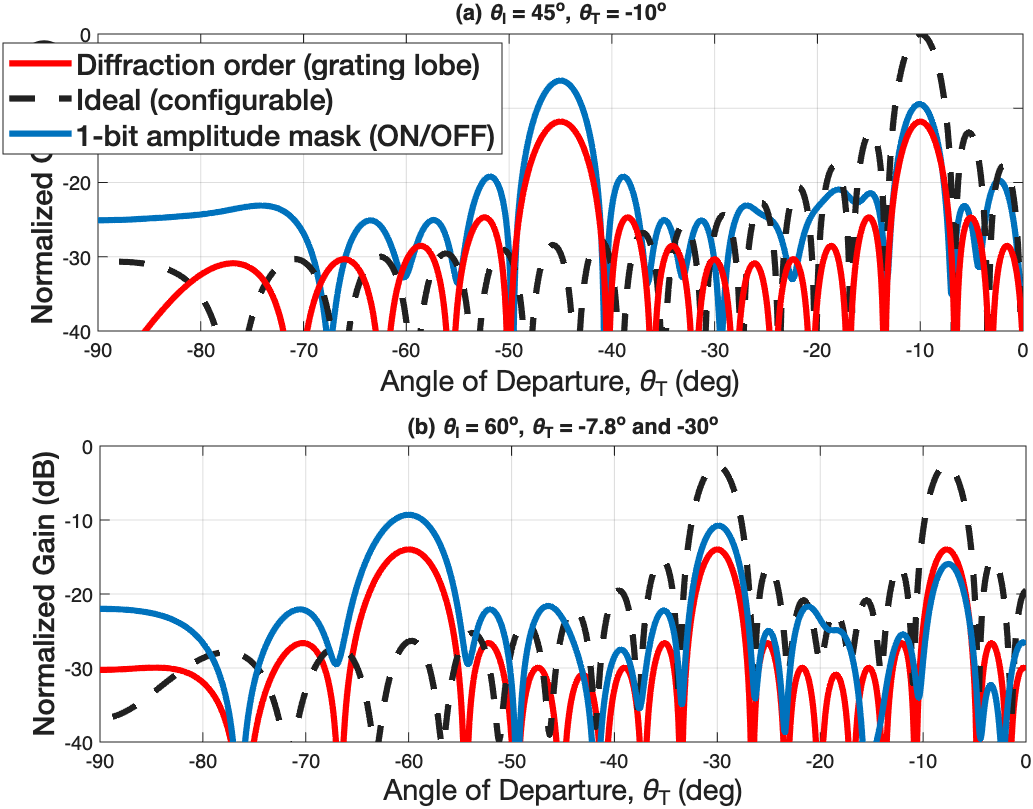}
\caption{Theory: aperture-matched comparison of diffraction-order steering and dense ON/OFF masking under fixed physical aperture constraints. (a) Single-target case for $\theta_I=45^\circ$ and $\theta_T=-10^\circ$. (b) Multi-target case (fixed footprint), illustrating reduced gain for diffraction-order designs when the period increase reduces effective sampling density.}
\label{fig:aperture_matched_DO_vs_binary}
\end{figure}

Diffraction-order designs can exhibit high target-lobe peaks in the unconstrained setting since selecting a larger
period $\delta$ increases the physical aperture when $M$ is held fixed. To enable a fair ``apples-to-apples''
comparison, Fig.~\ref{fig:aperture_matched_DO_vs_binary} evaluates diffraction-order  and dense ON/OFF masking under a fixed physical
footprint. In this aperture-matched setting, increasing $\delta$ reduces the
number of active cells or elements that fit within the fixed length and thereby lowers  the
coherent gain to the desired diffraction order.

As shown in Fig.~\ref{fig:aperture_matched_DO_vs_binary}(a), for the single-target case $(\theta_I,\theta_T)=(45^\circ,-10^\circ)$,
the dense ON/OFF mask on the $\lambda/2$ maintains a larger effective number of contributing cells and therefore
sustains stronger target response than the aperture-matched diffraction-order design. Fig.~\ref{fig:aperture_matched_DO_vs_binary}(b) shows
the same trend in the multi-target setting where periodic steering becomes less efficient due to a
sparser effective aperture.  This reduces coherent accumulation at each intended non-specular direction. 

  {
The associated sidelobe behavior follows from the underlying aperture structures.  In the ON/OFF case, binary amplitude quantization introduces additional spatial harmonics and does not fully suppress the residual specular component,
leading to stronger secondary lobes than in the ideal continuous-phase case. In the
diffraction-order case, additional lobes are an inherent consequence of the periodic
aperture structure, since the desired beam is realized as a selected diffraction order of
a passive periodic reflector.   Overall, these results indicate that under fixed-size deployment constraints, dense ON/OFF masking is typically
preferable because it preserves a larger effective aperture and stronger target-direction gain. By contrast,
diffraction-order steering is attractive when aperture expansion is permitted since its synthesis is simple
and analytically tractable. }

\subsubsection{  {Two-Dimensional Angular Response and Secondary-Lobe Behavior}}
\begin{figure}[t]
\hspace{-1mm}
\includegraphics[width=1.02\columnwidth]{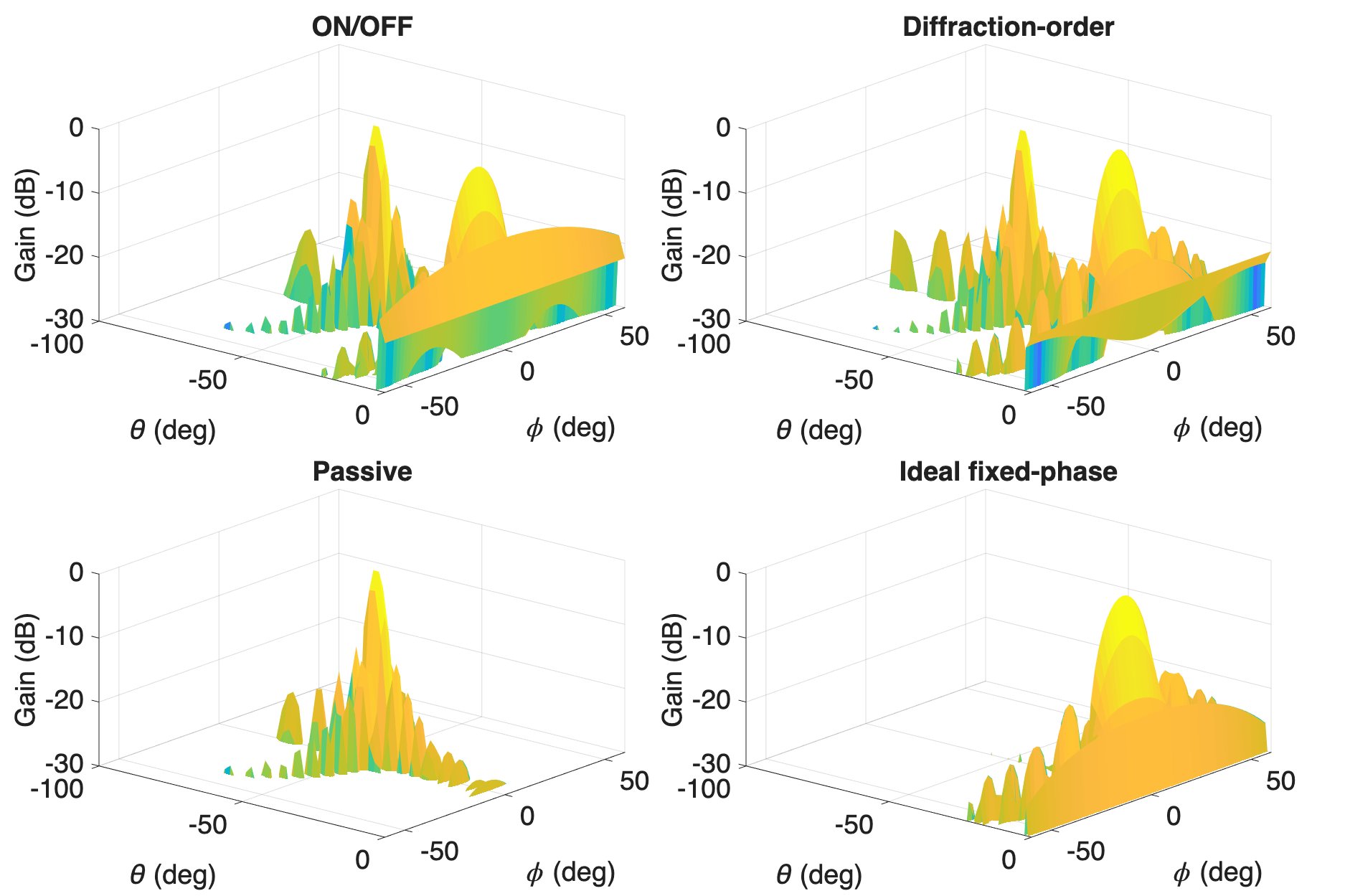}
\caption{  {Three-dimensional angular response patterns at 60~GHz for the representative single-beam case
$\theta_I=45^\circ$ and target $\theta_T=-10^\circ$. The responses are normalized to the peak of each
design at 60~GHz.}}
\label{fig:fig2_single}
\end{figure}

\begin{figure}[t]
\hspace{-1mm}
\includegraphics[width=1.02\columnwidth]{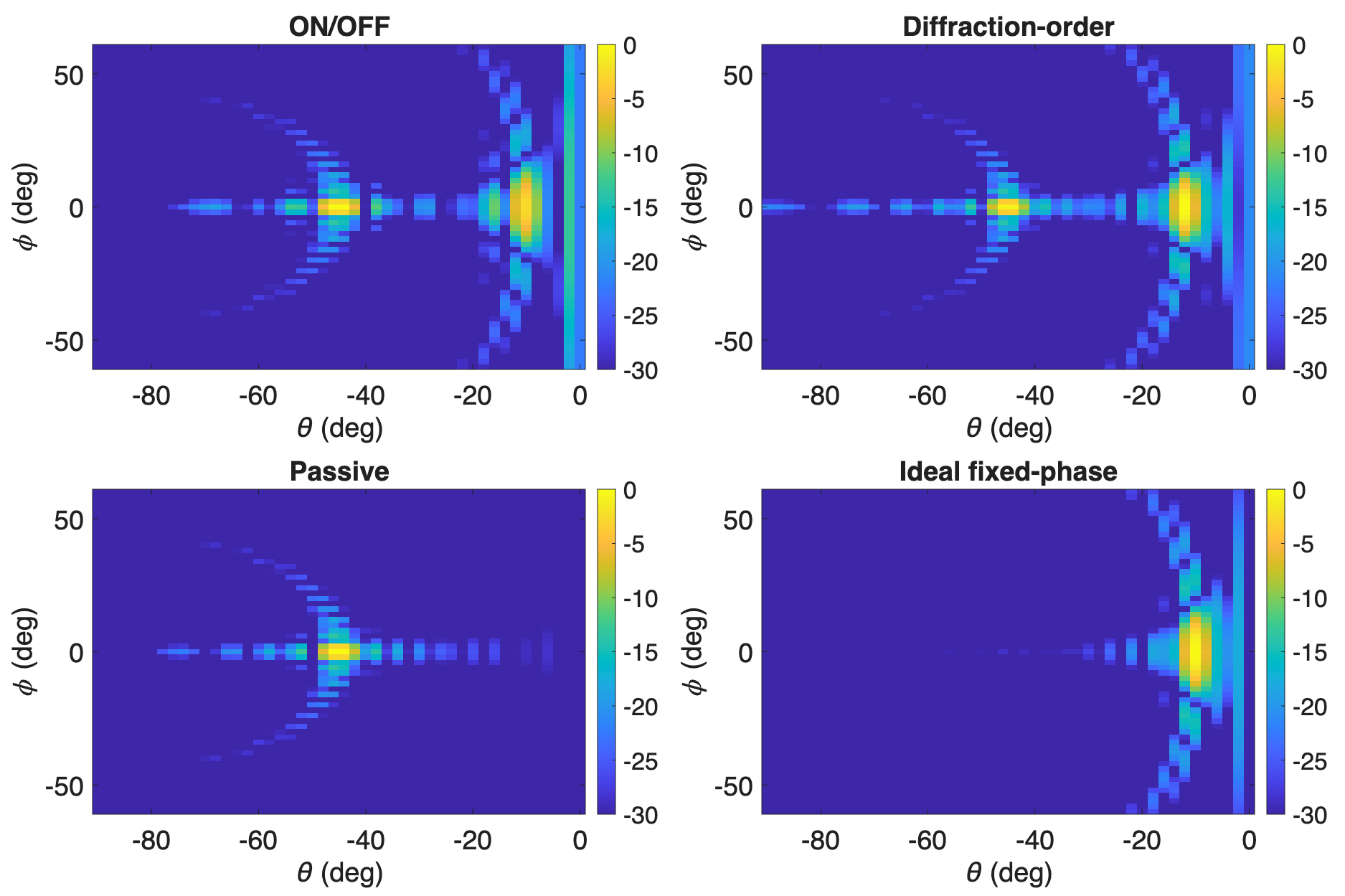}
\caption{  {Two-dimensional angular response maps at 60~GHz for the representative single-beam case
$\theta_I=45^\circ$ and target $\theta_T=-10^\circ$. The plots show the normalized response of the
proposed ON/OFF reflector, the diffraction-order reflector, the passive all-ON reflector, and the
ideal fixed-phase reflector, evaluated using the 2-D aperture model.}}
\label{fig:fig1_single}
\end{figure}

\begin{figure}[t]
\hspace{-1mm}
\includegraphics[width=1.02\columnwidth]{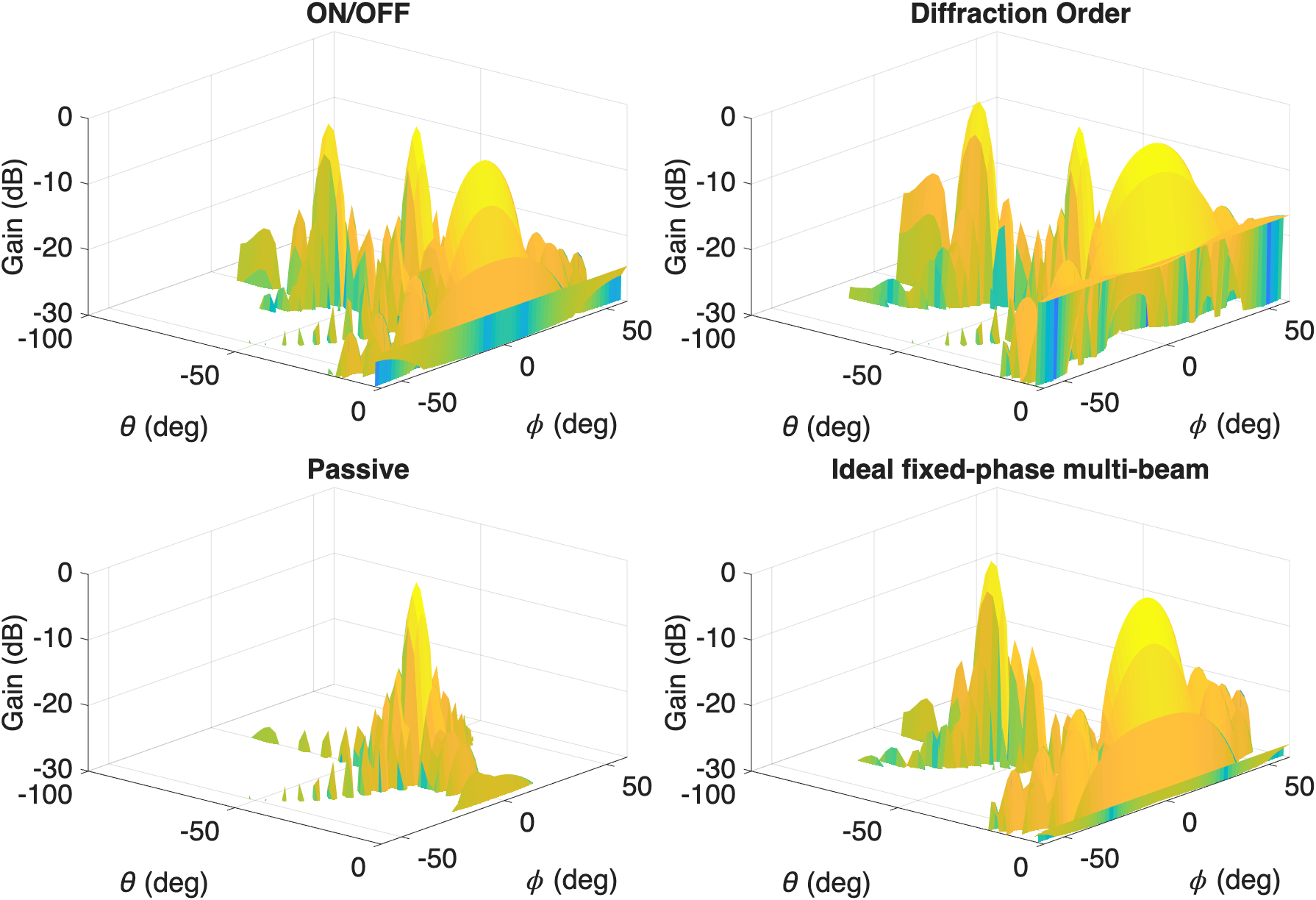}
\caption{  {Three-dimensional angular response patterns at 60~GHz for the representative multi-beam case
$\theta_I=30^\circ$ with target directions $\theta_T\in\{-7.8^\circ,-60^\circ\}$. }}
\label{fig:fig1_multi}
\end{figure}

\begin{figure}[t]
\hspace{-1mm}
\includegraphics[width=1.02\columnwidth]{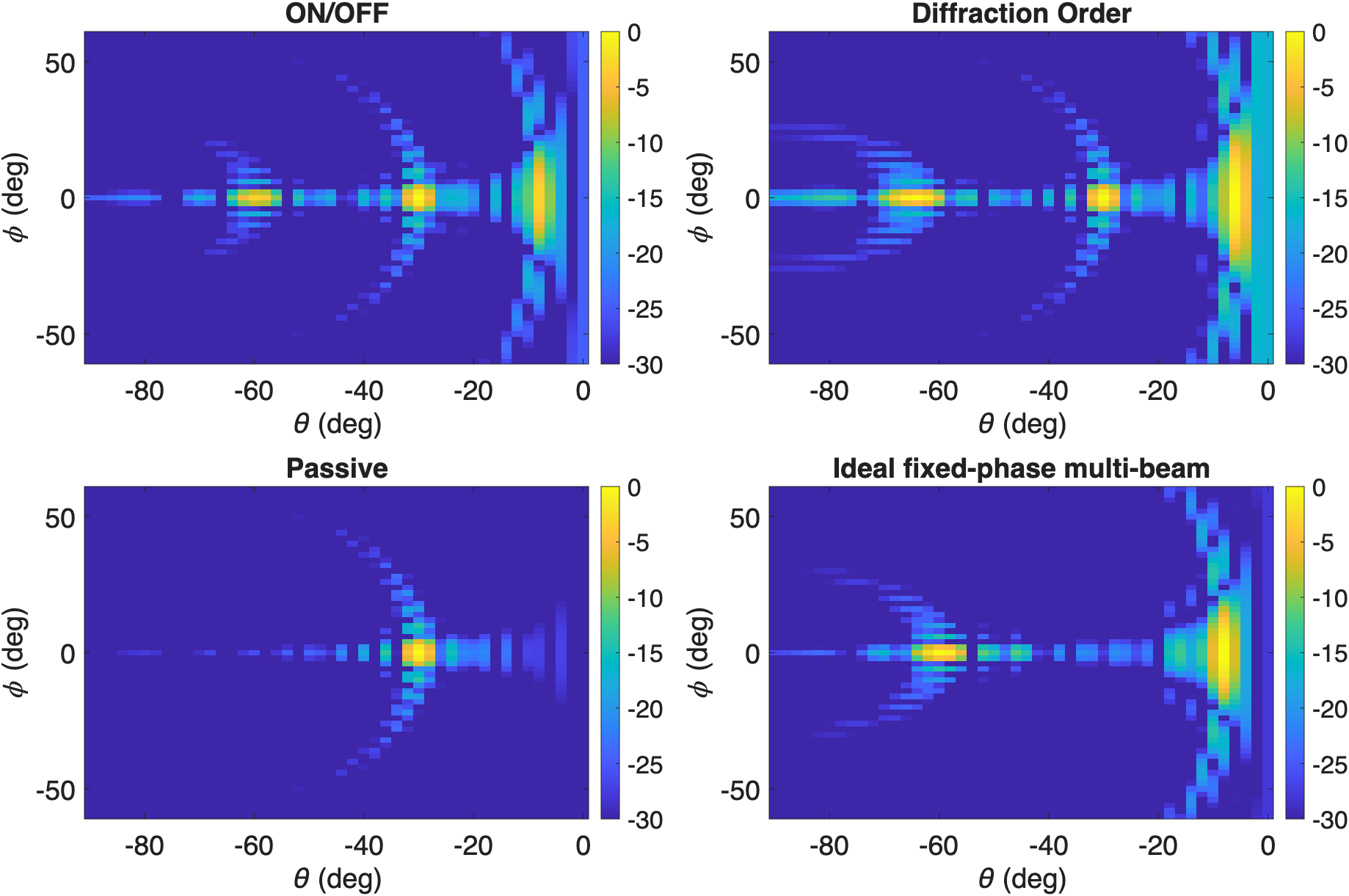}
\caption{  {Two-dimensional angular response maps at 60~GHz for the representative multi-beam case
$\theta_I=30^\circ$ with target directions $\theta_T\in\{-7.8^\circ,-60^\circ\}$. The plots show
the normalized response of the proposed ON/OFF reflector, a diffraction-order reflector,  the passive all-ON reflector, and the ideal fixed-phase
multi-beam reflector, evaluated using the 2-D aperture model.}}
 \label{fig:fig2_multi}
\end{figure}

  {To complement the azimuth-plane cuts, } Figs.~\ref{fig:fig2_single}--\ref{fig:fig2_multi}   {
show representative 3-D angular patterns and 2-D angular response maps obtained from the
planar-aperture model in} \eqref{eq:p_2d}   {for both the single-beam and multi-beam cases using a $35\times35$ lattice. 
These results are consistent with the 1-D azimuthal analysis, while also revealing the
response variation with the second angular coordinate $\phi$. Since the synthesis in this
paper is carried out for azimuth-plane steering, the strongest response remains centered
near $\phi=0$. As $|\phi|$ increases, the response decreases and broadens because the
design does not impose independent elevation steering.}

For the single-beam case, Figs.~\ref{fig:fig2_single} and \ref{fig:fig1_single}
correspond to $\theta_I=45^\circ$ and $\theta_T=-10^\circ$.   {The passive all-ON reflector
exhibits the expected specular response with dominant energy near
$\theta_{\mathrm{spec}}=-45^\circ$. The ideal fixed-phase reflector concentrates energy
near the desired target direction. The proposed ON/OFF design also forms a deterministic
non-specular lobe toward $\theta_T=-10^\circ$, but with broader secondary structure due to
binary quantization and residual specular leakage. The diffraction-order reflector likewise
produces a lobe near the target direction, but its periodic structure leads to more
pronounced secondary lobes.}

For the multi-beam case, Figs.~\ref{fig:fig1_multi} and \ref{fig:fig2_multi}
correspond to $\theta_I=30^\circ$ with target directions
$\theta_T\in\{-7.8^\circ,-60^\circ\}$.   {The passive all-ON reflector again remains
predominantly specular, whereas the ideal fixed-phase reflector produces two intended lobes
near the desired target directions. The ON/OFF reflector also enhances both target
directions, confirming that deterministic multi-beam shaping can be approximated with a
fully passive binary aperture. The diffraction-order reflector shown in these figures is a
single-target periodic reference and therefore does not realize both target beams
simultaneously. Overall, the 2-D aperture results reinforce the same main conclusion as the
1-D analysis. Fully passive binary-coded apertures can realize deterministic non-specular
beam shaping on a fixed dense lattice, while diffraction-order steering offers simpler
periodic synthesis at the cost of stronger secondary lobes and reduced flexibility for
multi-beam operation.}

\subsubsection{  {Wideband Sensitivity of Fixed Passive Designs}}
\begin{figure}[t]
\hspace{-1mm}
\includegraphics[width=1.02\columnwidth]{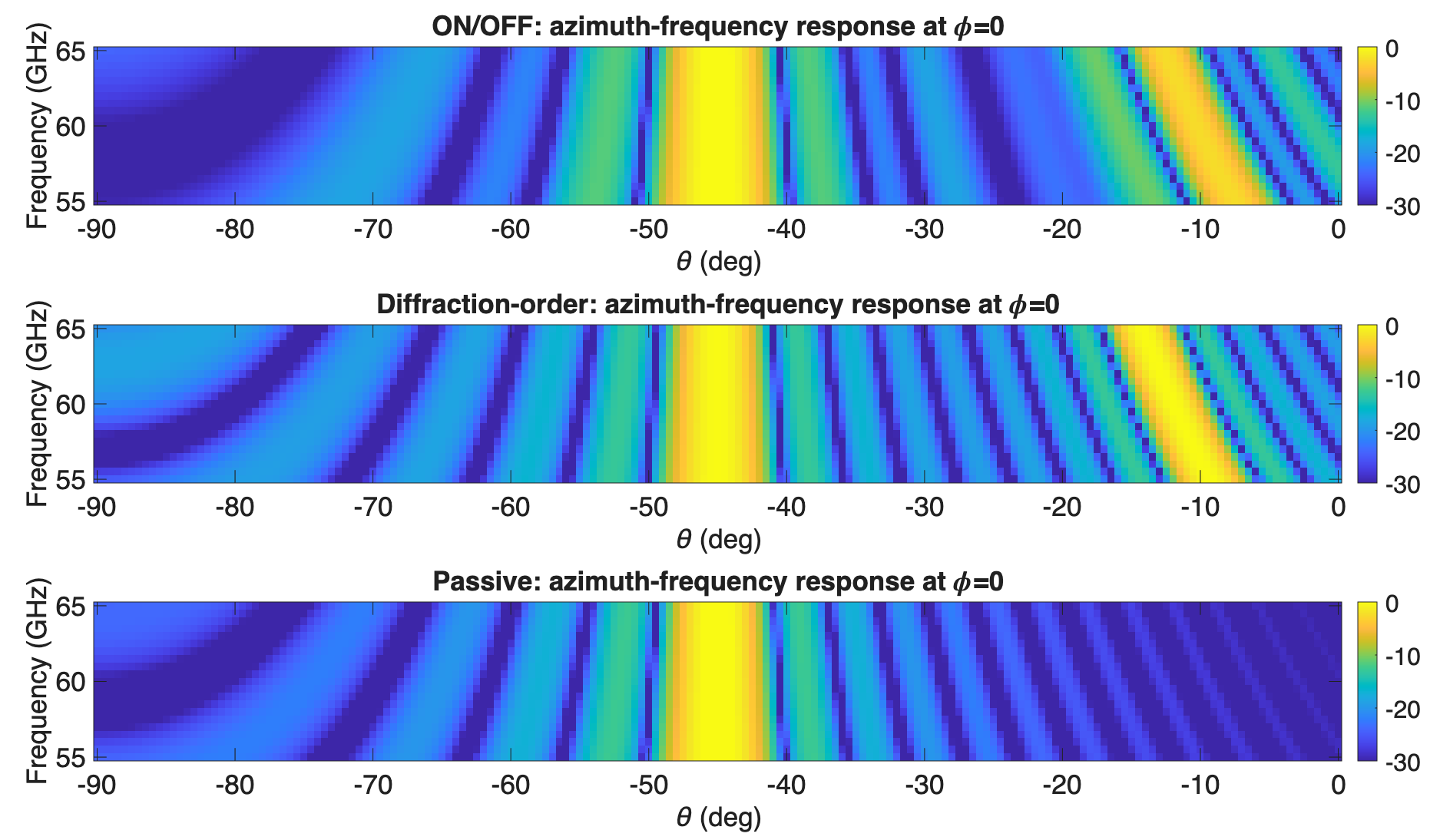}
\caption{  {Target-direction gain variation versus frequency for the single-beam case
$\theta_I=45^\circ$ and $\theta_T=-10^\circ$, normalized to the corresponding 60~GHz response of each
design.}}
\label{fig:fig7_single}
\end{figure}

\begin{figure}[t]
\hspace{-1mm}
\includegraphics[width=1.02\columnwidth]{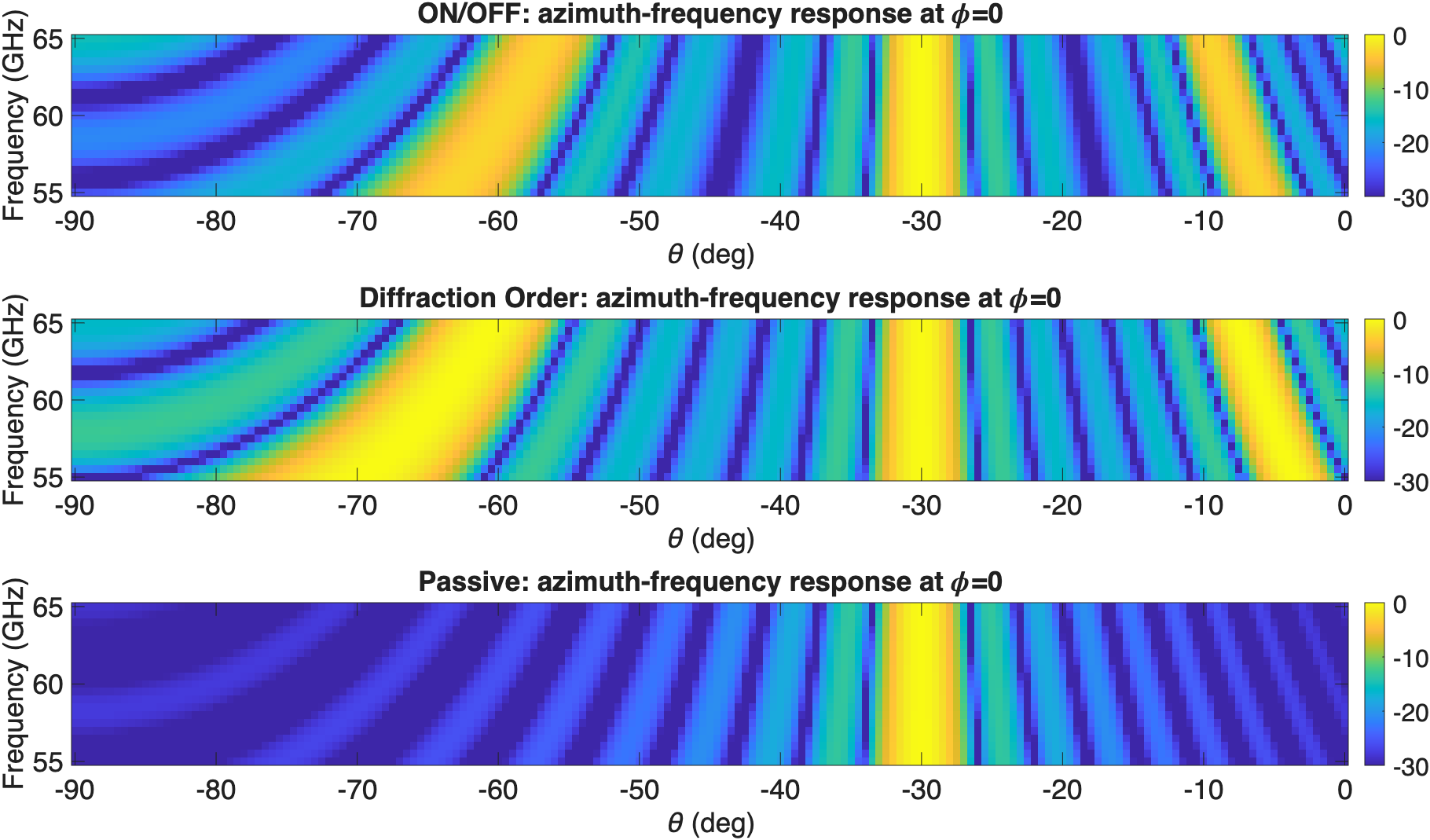}
\caption{  {Target-direction gain variation versus frequency for the multi-beam case
$\theta_I=30^\circ$ with target directions $\theta_T=-7.8^\circ$ and $\theta_T=-60^\circ$,
normalized to the corresponding 60~GHz response of each design. }}
\label{fig:fig7_multi}
\end{figure}

Figs.~\ref{fig:fig7_single} and \ref{fig:fig7_multi}   {quantify the wideband sensitivity of
the passive designs for the single-beam and multi-beam cases. In both cases, the
responses are normalized to the corresponding 60~GHz design-point value so that the
off-design loss can be observed directly as the carrier frequency is varied while the
physical reflector geometry is held fixed.

For the single-beam case in Fig.}~\ref{fig:fig7_single},   {all designs exhibit reduced
target-direction response as the operating frequency moves away from the design point. The
ON/OFF reflector shows a more gradual reduction in target-direction response than the
diffraction-order design. This behavior is expected because diffraction-order steering
depends explicitly on the ratio between the fixed spatial period and the wavelength, so the
selected diffraction order shifts more rapidly as frequency changes. By contrast, the dense
ON/OFF aperture is comparatively more tolerant to moderate off-design operation, although it
also experiences loss away from the design frequency.

A similar trend is observed in the multi-beam case shown in Fig.}~\ref{fig:fig7_multi}.    {The results
again show that fixed passive beam patterns are frequency-sensitive, with the
diffraction-order design exhibiting stronger degradation away from the 60~GHz design point.
Although the ON/OFF reflector is not frequency-invariant, it maintains more stable
target-direction behavior over moderate frequency variation and therefore provides a more
robust passive alternative under fixed-aperture operation.}

\subsubsection{Energy-efficiency implication under surface-power overhead}
\begin{figure}[t]
\centering
\includegraphics[width=0.80\columnwidth]{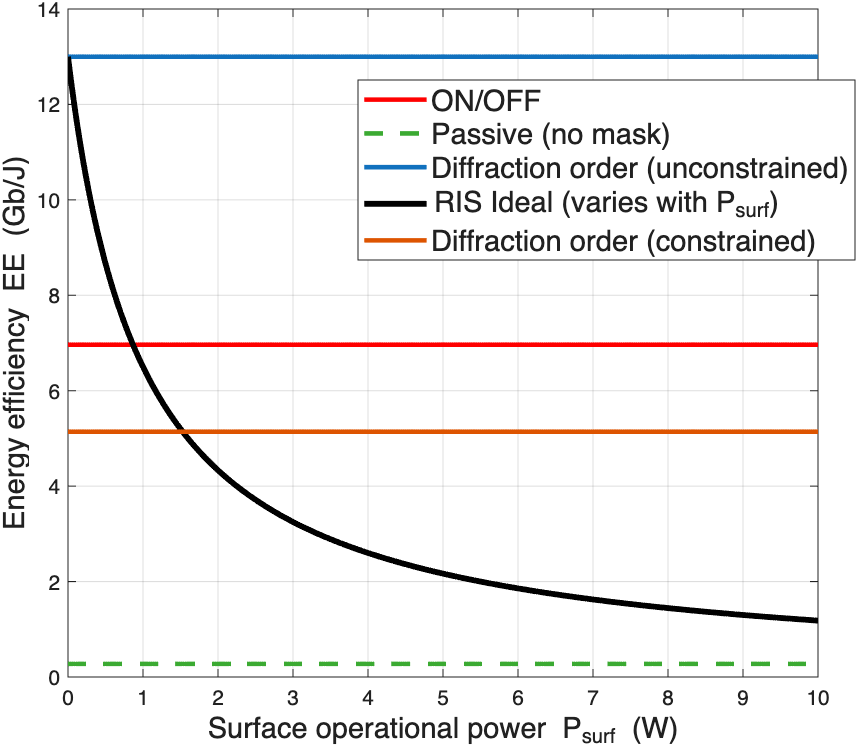}
\caption{Energy efficiency $\mathrm{EE}=R/(P_0+P_{\mathrm{surf}})$ versus surface operational power $P_{\mathrm{surf}}$ at $\theta_T=-10^\circ$ for an example setting with $P_0=1$~W, $B=2$~GHz, and baseline $\mathrm{SNR}_0=-10$~dB. Fully passive designs have $P_{\mathrm{surf}}=0$ and therefore yield constant $\mathrm{EE}$, whereas a powered RIS baseline decreases with $P_{\mathrm{surf}}$. For the diffraction-order curve, {unconstrained} indicates that the selected period is applied while allowing the physical aperture to expand.}
\label{fig:EE_vs_Psurf}
\end{figure}

Fig.~\ref{fig:EE_vs_Psurf} illustrates the impact of surface-side (reflector) operational power on energy efficiency at the
target direction $\theta_T=-10^\circ$. Using the definitions in Section~\ref{subsec:system_energy_metrics}, we evaluate EE
under a fixed radio-side power budget $P_0\triangleq P_{\mathrm{TX}}+P_{\mathrm{circ}}$ and vary only the surface-side term
$P_{\mathrm{surf}}$. The achievable rate is obtained from the target-direction SNR using \eqref{eq:rate_metric}. The example
parameters used to generate the curves are given in the figure caption.
For each reflector design, the target-direction $\mathrm{SNR}(\theta_T)$ is scaled according to its relative target gain, so
that the resulting curves capture the joint effect of reflector-induced SNR improvement and added surface-side power.

Since the proposed  reflectors are fully passive, $P_{\mathrm{surf}} = 0$ and their EE
curves remain essentially constant. In contrast, powered RIS baselines exhibit a monotonic EE reduction as $P_{\mathrm{surf}}$
increases, because controller/driver overhead and state-dependent unit-cell power increase the denominator of \eqref{eq:EE_metric}
without providing a commensurate increase in $R(\theta_T)$. This behavior is consistent with measurement-backed RIS power
decompositions in which the controller term alone can be several watts (e.g., $P_{\mathrm{ctrl}}\approx 4.8$~W \cite{WangTCOM2024Power,WangEUSIPCO2023StaticPower}). Overall,
Fig.~\ref{fig:EE_vs_Psurf} highlights that eliminating surface-side operational power allows target-direction beam gains to
translate more directly into higher bits/Joule, which is particularly relevant for static indoor installations.

\begin{table*}[t]
\centering
\caption{Background-subtracted (mount-only) relative received power for the 3D-printed inkwell reflectors in Fig.~\ref{fig:combined_prototypes} (\(M=35\) cells per side on the base scaffold prior to metallization; azimuth scan at fixed elevation).}
\label{tab:new_meas_bgs}
\renewcommand{\arraystretch}{1.10}
\begin{tabular}{@{} c c l c c c @{}}
\toprule
\textbf{AoA} & \textbf{AoD} & \textbf{Scheme} &
\(P_{\mathrm{bgs}}(\theta)\) (dB) &
\(\Delta P_{\mathrm{bgs}}(\theta)\) (dB) &
\(G_{\mathrm{norm}}(\theta)\) (dB) \\
\midrule

\multirow{5}{*}{\(45^\circ\)} & \multirow{5}{*}{\(-45^\circ\)}
&  Passive (all-ON; specular)             & \(-73.05\) & \(0.00\)  & \(0.00\) \\
& & Metal plate (baseline)                 & \(-72.93\) & \(+0.12\) & \(+0.12\) \\
& & Proposed ON/OFF (1-bit amplitude)      & \(-77.10\) & \(-4.05\) & \(-4.05\) \\
& & Diffraction-order (grating-lobe)       & \(-78.95\) & \(-5.90\) & \(-5.90\) \\
\midrule

\multirow{5}{*}{\(45^\circ\)} & \multirow{5}{*}{\(-10^\circ\)}
&  Passive (all-ON; specular)             & \(-96.57\) & \(0.00\)  & \(-23.52\) \\
& & Metal plate (baseline)                 & \(-95.67\) & \(+0.90\) & \(-22.62\) \\
& & Proposed ON/OFF (1-bit amplitude)      & \(-77.29\) & \(+19.27\) & \(-4.24\) \\
& & Diffraction-order (grating-lobe)       & \(-78.88\) & \(+17.69\) & \(-5.83\) \\
\midrule

\multirow{5}{*}{\(30^\circ\)} & \multirow{5}{*}{\(-7.8^\circ\)}
& Passive (all-ON; specular)             & \(-96.48\) & \(0.00\)  & \(-23.43\) \\
& & Metal plate (baseline)                 & \(-92.71\) & \(+3.77\) & \(-19.66\) \\
& & Proposed ON/OFF (1-bit amplitude)      & \(-78.70\) & \(+17.77\) & \(-5.65\) \\
& & Diffraction-order (grating-lobe)       & \(-79.79\) & \(+16.69\) & \(-6.74\) \\
\midrule

\multirow{5}{*}{\(30^\circ\)} & \multirow{5}{*}{\(-60^\circ\)}
&  Passive (all-ON; specular)             & \(-98.74\) & \(0.00\)  & \(-25.69\) \\
& & Metal plate (baseline)                 & \(-94.57\) & \(+4.16\) & \(-21.52\) \\
& & Proposed ON/OFF (1-bit amplitude)      & \(-79.18\) & \(+19.56\) & \(-6.13\) \\
& & Diffraction-order (grating-lobe)       & \(-81.41\) & \(+17.33\) & \(-8.36\) \\
\midrule

\multirow{5}{*}{\(30^\circ\)} & \multirow{5}{*}{\(-30^\circ\)}
&  Passive (all-ON; specular)             & \(-73.05\) & \(0.00\)  & \(0.00\) \\
& & Metal plate (baseline)                 & \(-72.89\) & \(+0.12\) & \(+0.12\) \\
& & Proposed ON/OFF (1-bit amplitude)      & \(-81.71\) & \(-8.66\) & \(-8.66\) \\
& & Diffraction-order (grating-lobe)       & \(-85.77\) & \(-12.72\) & \(-12.72\) \\

\bottomrule
\end{tabular}
\end{table*}
\subsection{Experimental Setup and Results}
\label{subsec:exp_results}

\subsubsection{Experimental setup and processing}
Over the air measurements are performed   {using a narrowband continuous-wave (CW) excitation at 60.48~GHz,
which serves as the design frequency for the fabricated reflectors shown in } Fig. \ref{fig:combined_prototypes}.   Two Sivers EVK02001 transceiver kits
are positioned on a constant-radius arc (0.9~m) centered on the reflector. The transmitter is fixed at
azimuth AoA $\theta_I$, while the receiver is swept in azimuth to measure the scattered response versus
AoD $\theta_T$. Two reference scans are collected, mount-only (background) and all-ON
(reference reflector). Background-subtracted power is computed as
\[
P_{\mathrm{bgs}}(\theta_T,\theta_I)=\max\{P_{\mathrm{meas}}(\theta_T,\theta_I)-P_{\mathrm{mount}}(\theta_T,\theta_I),\,0\},
\]
and target-direction results are reported using
\[
G_{\mathrm{norm}}(\theta_T)=10\log_{10}\!\left(\frac{P_{\mathrm{bgs}}(\theta_T)}{\max_{\theta}P_{\mathrm{bgs,all\text{-}on}}(\theta)}\right).
\]
  { Since the experimental measurements use a narrowband CW signal, they validate passive beam shaping at the design
frequency rather than broadband communication performance.}

\begin{figure}[t]
 \centering
 \includegraphics[width=0.35\textwidth]{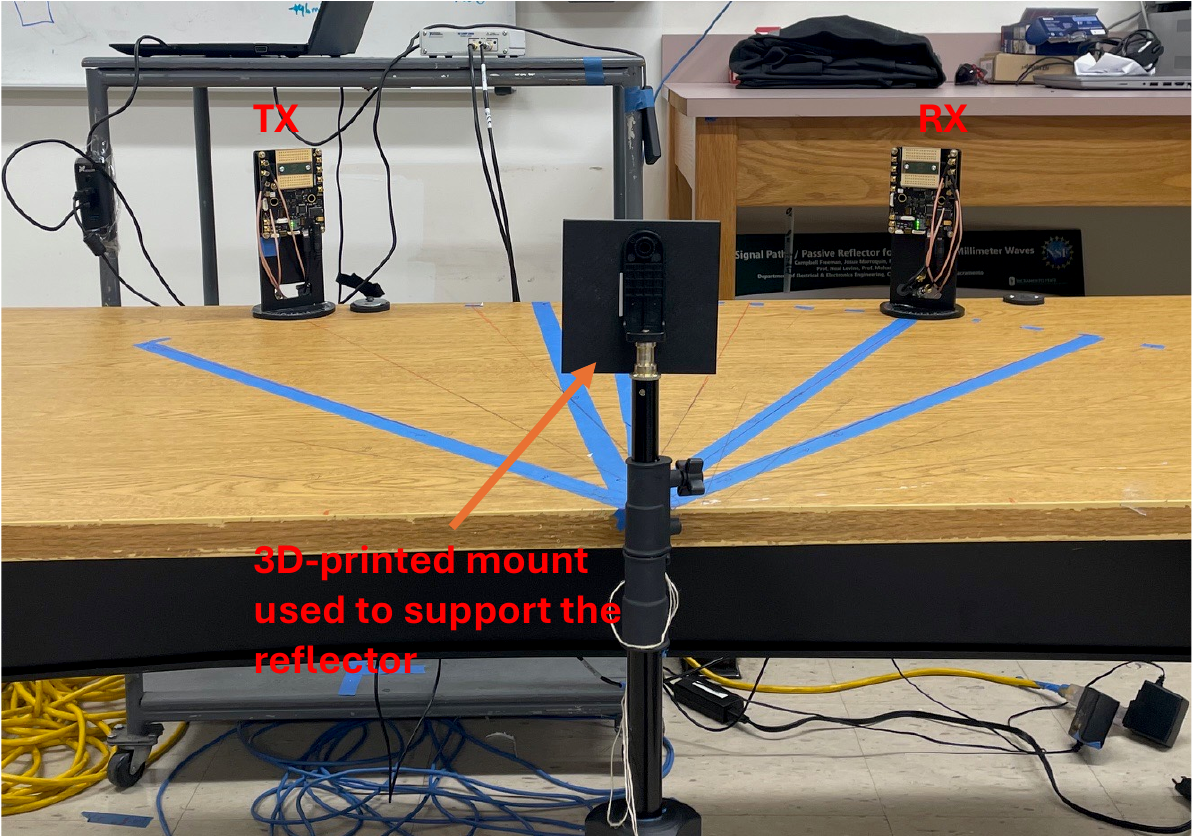}
 \caption{Over the air azimuthal sweep measurement setup. The transmitter (TX) is fixed at the prescribed incidence angle, while the receiver (RX) is swept in azimuth to measure the scattered response.}
 \label{fig:model1a}
\end{figure}

\subsubsection{Experimental Results and Discussion}
We experimentally evaluate the performance of the proposed passive reflector designs for both single-beam and multi-beam scenarios. The single-beam configuration considers an incidence angle of $\theta_I=45^\circ$ with a target direction $\theta_T=-10^\circ$, while the multi-beam case uses $\theta_I=30^\circ$ with target directions $\theta_T\in\{-7.8^\circ,-60^\circ\}$. In addition to the fabricated all-ON inkwell reflector, a smooth copper plate is included as a reference baseline to verify that the all-ON metallization achieves comparable specular reflection and to establish a consistent conductor benchmark for subsequent comparisons.

Table~\ref{tab:new_meas_bgs} reports the background-subtracted response $P_{\mathrm{bgs}}(\theta)$, the enhancement over the
passive all-ON baseline $\Delta P_{\mathrm{bgs}}(\theta)$, and the normalized gain $G_{\mathrm{norm}}(\theta)$. The metric
$G_{\mathrm{norm}}(\theta)$ measures the target-direction level relative to the {maximum} background-subtracted response
of the passive all-ON scan (specular lobe), and thus indicates how far a given angle lies below the strongest
all-ON direction. In contrast,
$\Delta P_{\mathrm{bgs}}(\theta)\!=\!P_{\mathrm{bgs,design}}(\theta)-P_{\mathrm{bgs,all\text{-}ON}}(\theta)$ directly
quantifies the gain improvement achieved by a given design over the passive all-ON reflector at the same $(\theta_I,\theta)$.

The measurements confirm that the fabricated all-ON reflector behaves similarly to a solid conductor. At the specular
directions, the copper-plate and all-ON responses are nearly identical, with only a small difference (about $0.1$~dB),
indicating that the inkwell metallization with the copper backing provides a high-reflectivity baseline prior to applying
any ON/OFF mask. Consistent with specular-dominated scattering under uniform weighting, the passive all-ON response is weak at
non-specular targets.  For $\theta_I=45^\circ$, the $-10^\circ$ direction is more than 20~dB below the all-ON peak in terms of
$G_{\mathrm{norm}}(\theta)$, demonstrating that an unshaped passive surface does not effectively redirect energy.

Applying ON/OFF masking and diffraction-order synthesis redistributes energy away from the specular direction and toward
the intended targets. This redistribution is visible as reduced specular response (negative $\Delta P_{\mathrm{bgs}}$ at the
specular angle) together with large positive enhancement at target angles. For the single-beam case
$(\theta_I,\theta_T)=(45^\circ,-10^\circ)$, the ON/OFF reflector provides a strong target enhancement
($\Delta P_{\mathrm{bgs}}\approx +19$~dB) and the diffraction-order design provides a comparable but slightly smaller improvement
($\approx +18$~dB) under the fixed-footprint constraint. These trends are consistent with the theoretical aperture-matched plots in
Fig.~\ref{fig:aperture_matched_DO_vs_binary}, where dense ON/OFF masking is favored when the physical aperture is fixed.
At the specular angle, both ON/OFF and diffraction-order incur a penalty relative to all-ON because the designs intentionally suppress or
reweight a portion of the aperture. For ON/OFF this is fundamentally tied to thinning (approximately $50\%$ activation ratio as shown in Lemma~\ref{lem:TT}).

For the multi-beam case ($\theta_I=30^\circ$), the ON/OFF and diffraction-order designs produce substantial enhancement at both target angles
($\theta_T=-7.8^\circ$ and $-60^\circ$), confirming successful deterministic non-specular beam formation in a fully passive
implementation. At the specular direction ($\theta_T=-30^\circ$), both designs exhibit reduced response relative to all-ON,
which is expected since energy is redirected toward the non-specular targets. We note here that the measured magnitude of these specular losses
and target gains do not always match the worst-case theoretical bounds relative to an {ideal} continuous-phase surface,
however, the experimentally relevant comparison is against the passive all-ON baseline, for which the measured target
enhancements are consistently large.  This is becuase the theory in Section~\ref{subsec:theory_results} is developed using a 1-D azimuthal array-factor abstraction, whereas the
fabricated reflector is a 2-D $35\times35$ lattice and the measurements are azimuth cuts of a 2-D scattering pattern. Additional
aperture extent in the orthogonal dimension, elevation-dependent weighting, and indoor multipath can affect absolute levels and
partially explain deviations from idealized 1-D predictions. Nevertheless, the measured ranking and trends across all-ON, ON/OFF,
and diffration order designs remain consistent with theory.

\begin{table}[t]
\caption{  {Measured response loss versus frequency for representative ON/OFF and diffraction-order
reflectors, normalized to the 60.48~GHz design point, for $\theta_I = 45^\circ$ and
$\theta_T = -10^\circ$.}}
\label{tab:response_loss}
\centering
\begin{tabular}{ccc}
\toprule
\textbf{Frequency (GHz)} & \textbf{ON/OFF loss (dB)} & \textbf{Diffraction-order loss (dB)} \\
\midrule
58.32 & -1.2  & -3.3  \\
60.48 & 0.0   & 0.0   \\
62.64 & -3.0  & -5.4  \\
64.80 & -8.0  & -11.5 \\
66.90 & -14.2 & -16.3 \\
69.10 & -13.4 & -20.7 \\
\bottomrule
\end{tabular}
\end{table}

  {
To experimentally assess frequency sensitivity beyond the nominal 60.48~GHz design point, additional narrowband CW measurements were conducted at 58.32, 60.48, 62.64, 64.80, 66.90, and 69.10~GHz for representative ON/OFF and diffraction-order reflectors in the beam-shaping configuration $\theta_I=45^\circ$ and $\theta_T=-10^\circ$. The measured responses were normalized to the 60.48~GHz result, and the corresponding losses are summarized in Table~}\ref{tab:response_loss}.    {The results show that the target-direction response at $\theta_T=-10^\circ$ degrades for both reflector types as the operating frequency shifts away from the design frequency. Across all measured frequencies, the diffraction-order reflector consistently exhibits larger loss than the ON/OFF design. For example, at 64.80~GHz the ON/OFF reflector experiences a loss of $-8.0$~dB compared to $-11.5$~dB for the diffraction-order design, while at 69.10~GHz the losses increase to $-13.4$~dB and $-20.7$~dB, respectively. This behavior is in close agreement with the analytical trends in Fig.~}\ref{fig:fig7_single},   {confirming that the ON/OFF design is comparatively more robust to frequency variation.  This degradation can be understood in terms of the wavelength dependence of the underlying beamforming mechanisms.  Diffraction-order steering relies explicitly on the ratio between the fixed spatial period and the wavelength (see} (\ref{eq:delta_star_1})),    {such that changes in frequency shift the effective diffraction condition and reduce gain in the intended direction. In contrast, the ON/OFF design relies on dense spatial modulation of the aperture and is therefore less sensitive to moderate changes in wavelength, although it still experiences performance degradation away from the design point. }
\section{Conclusion}
\label{sec:conclusion}
This paper presented a theory-to-hardware framework for fully passive millimeter-wave beam shaping using in-lab fabricated, low-cost 3D-printed reflectors. Beam patterns are fabrication-coded on a fixed dense lattice using binary masks or periodic activation and realized via stencil-assisted conductive deposition on a copper-backed ``inkwell'' substrate. Two complementary passive mechanisms were developed and validated: fixed-aperture 1-bit ON/OFF spatial masking for non-specular steering and multi-beam synthesis, and diffraction-order (grating-lobe) steering via uniform period selection to place a chosen diffraction order at a desired departure angle.

The experimental results highlight an important tradeoff between these mechanisms. Diffraction-order steering is attractive due to its simplicity, but it typically produces narrower angular features and is therefore more sensitive to alignment errors and geometric mismatch. It is also more susceptible to beam squint under wideband operation because the grating condition is wavelength dependent.  This reduces coherence at a fixed target direction as frequency or angle deviates from the design point. In contrast, dense ON/OFF masking operates on a fixed footprint and provides greater flexibility for shaping the angular response (including multi-beam patterns), with improved robustness to practical deployment tolerances.

Future work will extend the framework to full two-dimensional aperture synthesis for joint azimuth--elevation beam control and polarization engineering, and will investigate semi-static control mechanisms (e.g., swappable masks or mechanically indexed patterns) suitable for indoor millimeter-wave electromagnetic environment shaping. We will also study pattern improvements through fabrication and material optimization, including the impact of cell width (effective metallized aperture per cell), metallization thickness/uniformity, and deposition methods to reduce ohmic loss and improve coherent reradiation efficiency.

\appendices
\section{Proof of Lemma~\ref{lem:TT}: Asymptotic $50\%$ Activation Ratio of the Cosine-Threshold Mask}
\label{app:proof_lemma1}
Recall that the element locations are $
x_m=\Bigl(m-\frac{M-1}{2}\Bigr)d_0,$
and define the phase associated with the $m$th element as
\[
\phi_m \triangleq kx_m(\sin\theta_T+\sin\theta_I),
\qquad
k=\frac{2\pi}{\lambda}.
\]
Substituting $x_m$ into $\phi_m$ yields the affine phase progression
\begin{align}
\phi_m
&= k\Bigl(m-\frac{M-1}{2}\Bigr)d_0(\sin\theta_T+\sin\theta_I) \nonumber\\
&= \Delta m + \beta_M,
\label{eq:phi_affine_app}
\end{align}
where
$
\Delta \triangleq k d_0(\sin\theta_T+\sin\theta_I),
\qquad
\beta_M \triangleq -\Delta\frac{M-1}{2}.
$
Hence, the phase increment is constant and satisfies $\phi_{m+1}-\phi_m=\Delta$. The cosine-threshold activation rule is given by
\[
b_m=\mathbf{1}\{\cos(\phi_m)\ge 0\}.
\]
Since $\cos(\cdot)$ is $2\pi$-periodic, the activation depends only on
$\phi_m \bmod 2\pi$. Introduce the normalized phase variable
\[
u_m \triangleq \frac{\phi_m}{2\pi}\bmod 1
= \Bigl(\frac{\Delta}{2\pi}m + \frac{\beta_M}{2\pi}\Bigr)\bmod 1.
\]
Under the condition $\Delta/(2\pi)\notin\mathbb{Q}$ (equivalently
$\Delta/\pi\notin\mathbb{Q}$), the sequence $\{u_m\}_{m\ge 0}$ is uniformly
distributed on $[0,1)$ by Weyl’s equidistribution theorem. Equivalently, $\phi_m \bmod 2\pi$ is uniformly distributed on the unit circle, with $u_m$ providing a convenient normalized representation.
To relate equidistribution to the cosine-threshold activation rule, define the
function
\[
f(u)\triangleq \mathbf{1}\{\cos(2\pi u)\ge 0\}.
\]
This choice of $f$ directly encodes the ON/OFF masking rule in normalized phase
coordinates, since
\[
b_m=\mathbf{1}\{\cos(\phi_m)\ge 0\}
=\mathbf{1}\{\cos(2\pi u_m)\ge 0\}
=f(u_m).
\]
By the defining consequence of equidistribution, for any Riemann-integrable
function $f:[0,1)\to\mathbb{R}$,
\[
\frac{1}{M}\sum_{m=0}^{M-1} f(u_m)
\;\xrightarrow[M\to\infty]{}\;
\int_0^1 f(u)\,du.
\]
Applying this result to the above choice of $f(u)$ yields
\begin{align}
\eta_M
&=\frac{1}{M}\sum_{m=0}^{M-1}\mathbf{1}\{\cos(\phi_m)\ge 0\} \nonumber\\
&=\frac{1}{M}\sum_{m=0}^{M-1} f(u_m)
\xrightarrow[M\to\infty]{}\;
\int_0^1 \mathbf{1}\{\cos(2\pi u)\ge 0\}\,du.
\end{align}
The limiting integral corresponds to the fraction of the unit interval for which
the cosine-threshold condition is satisfied. Define the set
\[
\mathcal{S}_+ \triangleq
\{u\in[0,1):\cos(2\pi u)\ge 0\}
=
\Bigl[0,\tfrac14\Bigr]\cup\Bigl[\tfrac34,1\Bigr).
\]
Since the integral of an indicator function equals the Lebesgue measure of its
support, we obtain
\[
\int_0^1 \mathbf{1}\{\cos(2\pi u)\ge 0\}\,du
= |\mathcal{S}_+|
= \frac{1}{2}.
\]
Therefore,
\[
\eta_M \xrightarrow[M\to\infty]{} \frac{1}{2}.
\]
This establishes the asymptotic $50\%$ activation ratio and completes the proof
of Lemma~\ref{lem:TT}.

\section{Proof of Lemma~\ref{lem:onoff_lb}}
\label{app:onoff_proof}

Consider a complex number $z$,  its magnitude can be written as $ |z|=\max_{\varphi\in[0,2\pi)}\Re\{e^{-j\varphi}z\}.
$
Applying this to $S_M^\star$ in (\ref{eq:coherent_sum_onoff}) and exchanging the two maximizations gives
\begin{align}
S_M^\star
&= \max_{\mathbf b\in\{0,1\}^M}\max_{\varphi\in[0,2\pi)} \Re\!\Big\{e^{-j\varphi}\sum_{m=0}^{M-1} b_m e^{-j\phi_m}\Big\} \nonumber\\
&= \max_{\varphi\in[0,2\pi)}\max_{\mathbf b\in\{0,1\}^M}
\sum_{m=0}^{M-1} b_m \cos(\phi_m+\varphi).
\label{eq:swapmax_app_varphi}
\end{align}
For fixed $\varphi$, the inner maximization is separable in $m$ and, since
$b_m\in\{0,1\}$, we have
$
\max_{b_m\in\{0,1\}} b_m\cos(\phi_m+\varphi) = [\cos(\phi_m+\varphi)]_+,
$
where $[x]_+\triangleq\max\{x,0\}$.  Define the normalized objective
$
F_M(\varphi)\triangleq \frac{1}{M}\sum_{m=0}^{M-1}[\cos(\phi_m+\varphi)]_+,
$
so that \eqref{eq:swapmax_app_varphi} can be written compactly as
\begin{equation}
\label{eq:Sstar_F_app_varphi}
\frac{S_M^\star}{M} = \max_{\varphi\in[0,2\pi)} F_M(\varphi).
\end{equation}
A distribution-free lower bound can be obtained by noting that the maximum of a function is no smaller than its average, hence
\begin{equation}
\label{eq:max_ge_avg_app_varphi}
\max_{\varphi} F_M(\varphi) \ge \frac{1}{2\pi}\int_0^{2\pi} F_M(\varphi)\,d\varphi.
\end{equation}
Interchanging the finite sum and the integral (by linearity) and then applying the
change of variables $u=\phi_m+\varphi$ gives
\begin{align}
\frac{1}{2\pi}\int_0^{2\pi} F_M(\varphi)\,d\varphi
&= \frac{1}{M}\sum_{m=0}^{M-1}\frac{1}{2\pi}\int_0^{2\pi}[\cos(\phi_m+\varphi)]_+\,d\varphi \nonumber\\
&= \frac{1}{2\pi}\int_0^{2\pi}[\cos u]_+\,du,
\label{eq:avg_reduce_app_varphi}
\end{align}
where the first equality follows from linearity and the change of variables
$u=\phi_m+\varphi$ is used in the second step. Since $[\cos(\cdot)]_+$ is
$2\pi$-periodic, each term in the sum is identical and independent of $m$.
Averaging therefore removes
element-dependent phase offsets.  Note the integrand $[\cos u]_+$ equals $\cos u$ on $u\in[-\tfrac{\pi}{2},\tfrac{\pi}{2}]$ (mod $2\pi$) and is zero elsewhere; therefore
$
\int_0^{2\pi}[\cos u]_+\,du=\int_{-\pi/2}^{\pi/2}\cos u\,du = 2.
$
Combining with \eqref{eq:avg_reduce_app_varphi} yields
$
\frac{1}{2\pi}\int_0^{2\pi} F_M(\varphi)\,d\varphi = \frac{1}{\pi}.
$
Using \eqref{eq:Sstar_F_app_varphi} and \eqref{eq:max_ge_avg_app_varphi} we conclude
\[
\frac{S_M^\star}{M} = \max_{\varphi}F_M(\varphi) \ge \frac{1}{\pi},
\]
which proves the amplitude bound \eqref{eq:onoff_amp_lb}. Squaring both sides yields
\(\gamma^\star \ge 1/\pi^2\), completing the proof.

\end{document}